\documentclass[twocolumn,prd,showpacs,floats,letterpaper,groupedaddress,superscriptaddress,amsmath,amsfonts,eqsecnum]{revtex4}

\usepackage{bm,graphics,graphicx,epsfig,color}

\newcommand{\Msun}{M_{\odot}}
\newcommand{\Mc}{{\cal M}}

\newcommand{\Inspnest}{\emph{Inspnest}~}
\newcommand{\SPINspiral}{\emph{SPINspiral}~}
\newcommand{\LALInferenceMCMC}{\emph{LALInferenceMCMC}~}

\newcommand{\be}{\begin{equation}}
\newcommand{\ee}{\end{equation}}
\newcommand{\bel}[1]{\begin{equation}\label{#1}}
\newcommand{\ba}{\begin{eqnarray}}
\newcommand{\ea}{\end{eqnarray}}
\newcommand{\bal}[1]{\begin{eqnarray}\label{#1}}

\begin{document}

\title{Estimating parameters of coalescing compact binaries with proposed advanced detector networks}
%\title{Estimating parameters of coalescing compact binaries with a detector network including LIGO Australia} 
%The effects of a detector network including LIGO Australia on compact binary parameter estimation}
\author{J. Veitch} \email{john.veitch@astro.cf.ac.uk}
\affiliation{School of Physics and Astronomy, Cardiff University, 5, The Parade, Cardiff, UK, CF24 3YB}
\author{I. Mandel} \email{imandel@star.sr.bham.ac.uk}
\affiliation{NSF Astronomy and Astrophysics Postdoctoral Fellow, MIT Kavli Institute, Cambridge, MA 02139} 
\affiliation{School of Physics and Astronomy, University of Birmingham, Edgbaston, Birmingham, B15 2TT}
\author{B. Aylott} 
\affiliation{School of Physics and Astronomy, University of Birmingham, Edgbaston, Birmingham, B15 2TT}
\author{B. Farr} 
\affiliation{Center for Interdisciplinary Exploration and Research in Astrophysics (CIERA) \& Dept.~of Physics and Astronomy, Northwestern University, 2145 Sheridan Rd, Evanston, IL 60208, USA
}
\author{V. Raymond}
\author{C. Rodriguez}
\affiliation{Center for Interdisciplinary Exploration and Research in Astrophysics (CIERA) \& Dept.~of Physics and Astronomy, Northwestern University, 2145 Sheridan Rd, Evanston, IL 60208, USA
}
\author{M. van der Sluys}  
\affiliation{Radboud University Nijmegen, P.O. Box 9010, NL-6500 GL Nijmegen, The Netherlands}
\author{V. Kalogera}
\affiliation{Center for Interdisciplinary Exploration and Research in Astrophysics (CIERA) \& Dept.~of Physics and Astronomy, Northwestern University, 2145 Sheridan Rd, Evanston, IL 60208, USA
}
\author{A. Vecchio} 
\affiliation{School of Physics and Astronomy, University of Birmingham, Edgbaston, Birmingham, B15 2TT}

\date{\today}

\begin{abstract}

%One of the goals of gravitational-wave astronomy is simultaneous detection of gravitational-wave signals from merging compact-object binaries and the electromagnetic transients from these mergers.
The advanced versions of the LIGO and Virgo ground-based gravitational-wave detectors are expected to operate from three sites: Hanford, Livingston, and Cascina.  Recent proposals have been made to place a fourth site in Australia or India; and there is the possibility of using the Large Cryogenic Gravitational Wave Telescope in Japan to further extend the network.
%With the next generation of advanced ground-based gravitational wave detectors under construction, we examine the benefits of the proposed extension of the detector network to include a fourth site in Australia in addition to the network of Hanford, Livingston and Cascina sites.
Using Bayesian parameter-estimation analyses of simulated gravitational-wave signals from a range of
coalescing-binary locations and orientations at fixed distance or signal-to-noise ratio, 
we study the improvement in parameter estimation for the proposed networks.  We find that a fourth detector site can break degeneracies in several parameters; in particular, the localization of the source on the sky is improved by a factor of $\sim 3$--$4$ for an Australian site, or $\sim 2.5$--$3.5$ for an Indian site, with more modest improvements in distance and binary inclination estimates.  This enhanced ability to localize sources on the sky will be crucial in any search for electromagnetic counterparts to detected gravitational-wave signals.

\end{abstract}

\preprint{{LIGO-P1100066}}

\maketitle

\section{Introduction}\label{sec:Introduction}

% LIGO History, upcoming Advanced LIGO, references, etc etc

Gravitational waves and electromagnetic counterparts from the merger of compact binaries will carry complementary information, and the successful association of the two types of merger signatures will allow many crucial questions in stellar and binary evolution and cosmology to be answered (see \cite{Bloom:2009,NuttallSutton:2010} and references therein).  Good sky localization of gravitational-wave sources will be crucial in searching for associated electromagnetic transients.  In this paper, we discuss ways in which relocating one of the LIGO detectors to a site in Australia or India could improve the prospects of multi-messenger gravitational-wave astronomy.

The final S6 science run of the enhanced-LIGO gravitational wave detectors~\cite{LIGO}, along with the third science run of the Virgo detector~\cite{Virgo} and GEO 600~\cite{GEO600}, has recently concluded.  Construction of the second generation of instruments is already underway, with 4\,km Advanced LIGO detectors undergoing installation at the Hanford, WA and Livingston, LA observatories~\cite{AdLIGO}, with sensitivity expected to improve by about one order of magnitude.
These two sites in North America are expected to be joined by the Advanced Virgo \cite{AdVirgo} detector, located in Cascina, Italy, to form a second-generation network consisting of three sites.
Recently, proposals have been made to add a fourth site to the network, at Gingin in Western Australia (- $31^\circ21^\prime28^{\prime\prime}$~S, $115^\circ42^\prime50^{\prime\prime}$~E) ,  or outside Bangalore in India ($14^\circ14^\prime$~N, $76^\circ26^\prime$~E).
The advantages and disadvantages of the LIGO Australia and LIGO India proposals were studied by a working group within the LIGO Scientific Collaboration, and their conclusions for LIGO Australia reported in~\cite{LIGOAustraliaReport}.
The Large Cryogenic Gravitational Wave Telescope (LCGT) detector is a planned interferometer with 3\,km arms and cryogenically cooled mirrors, located in Japan ($36^\circ15^\prime$~N, $137^\circ11^\prime$~E) 200 metres below ground to reduce seismic noise~\cite{Kuroda:LCGT}. In addition to the 4-site advanced LIGO/Virgo networks, we also consider the possible network with two advanced LIGO detectors at Hanford, one at Livingston; advanced Virgo at Cascina and LCGT in Japan. In the following, we use an acronym for denoting the network based on the first initials of the sites involved: Australia (A), Hanford (H), India (I), Japan (J), Livingston (L) and Virgo (V); e.g., AHLV is a network consisting of LIGO detectors in Australia, Hanford and Livingston and the Virgo detector.

In this study, we make a comparison of the performance of the proposed networks with the performance of the 3-site HHLV network in terms of parameter estimation for compact binaries.  We focus on binary neutron-star systems, which are expected to be a prevalent source of observable signals for the advanced detector network~\cite{RatesPaper}.

Using independent Bayesian analyses, we compare the parameter estimation performance of each network for an ensemble of sources spread throughout parameter space, and demonstrate the improvement gained from the addition of sites to the network. The relative improvement for individual sources is assessed at a fixed distance or signal-to-noise ratio. Bayesian inference allows us to extract all of the available information about system parameters from the full data set.  In contrast with the timing triangulation method discussed below, all of the data from all interferometers are taken into account coherently. And unlike Fisher information matrix techniques, which explicitly only describe the local structure of the parameter space and are accurate only in the limit of high signal-to-noise ratio (SNR), Bayesian inference methods search through the full parameter space, showing the global structure of posteriors, including multiple modes or near-degeneracies.

Beginning with the work of \cite{Jaranowski:1996}, several other studies have analyzed sky-localization accuracy with different network configurations \cite{FairhurstTiming,FairhurstAdvancedNetworks,Schutz:2011,WenChen:2010}.  These investigations either used only a limited amount of information (e.g., timing information alone) or Fisher-matrix techniques that may fail in a multimodal, degenerate parameter space.  The study by Nissanke \emph{et al.} \cite{Nissanke:2011} used Bayesian methods to compare HLV, AHLV, HJLV and AHJLV networks, but used a single detector at Hanford for the HLV case (so some of their quoted improvements are simply due to greater SNR accessible with 4 rather than 3 detectors). Furthermore, \cite{Nissanke:2011} used a population of sources randomly distributed in space out to $z=1$, whereas we distributed our sources at constant distance (for nested sampling [NS] and Markov Chain Monte Carlo [MCMC] methods) or constant SNR (Fisher matrix method) and varied the other extrinsic parameters. This was done to ensure good coverage of the extrinsic parameter space when assessing the relative network performances. Meanwhile, the study by Klimenko \emph{et al.} focused on sky localization of transient burst sources rather than coalescing compact binaries with known waveforms \cite{Klimenko:2011}.

The ``Report of the Committee to Compare the Scientific Cases for AHLV and HHLV'', which considered the scientific advantages of moving one of the Hanford detectors to Australia \cite{LIGOAustraliaReport}, was based in part on a preliminary version of the work presented here.  Ours is the first study to apply Bayesian techniques to comparisons of parameter-estimation accuracies with a network including an Indian detector as well as networks with Australian or Japanese detectors.  Furthermore, we analyze the impact of moving one of the detectors to Australia or India on the accuracy of measuring masses, distances, and other parameters in addition to sky localization for the first time~\footnote{The impact of an Australian detector on distance and inclination measurements was discussed in \cite{Nissanke:2009} in a very different context, where the sky location was assumed to be known perfectly because of a presumed detection of an electromagnetic counterpart.}.

This paper is organized as follows.  In section II, we describe the network configurations being analyzed.  In section III, we introduce the analysis techniques employed in this paper.  The simulations and their results are described in section IV.  We conclude in section V.  We also include an appendix discussing the impact of the change of network configurations on false alarm probabilities and detection thresholds for a fully coherent analysis.

% Description of networks.
\section{Network configurations}

Interferometric gravitational wave detectors are notoriously bad at determining the direction of incoming radiation from short duration sources when used individually, as the detector has good sensitivity over a large range of angles. Although this allows an all-sky search to be performed without the need for `pointing', it also means that the amplitude of the incoming signal cannot be used to determine its location well. To be able to resolve the position on the sky of a short duration gravitational wave source, a network of interferometers is needed. 
%The differences in the time of arrival of the gravitational wave source at the different sites can then be used to perform triangulation of the source location.

With more than one site it is possible to use triangulation to determine the location of an observed signal using the observed time delay between different detectors. With two sites, this method can resolve the position to within a ring on the sky centered on the axes between the sites; with three sites this is reduced to two patches; but the addition of a fourth site allows a unique patch to be determined for each source. The limiting factors in the accuracy of the timing method are the distances between the sites in the network and the timing accuracy of the sites, which is itself governed by the signal-to-noise ratio and the effective bandwidth of the signal in each detector~\cite{FairhurstTiming}. So, to achieve a better sky resolution for any particular gravitational wave, a network should consist of detectors with a wide band of sensitivity, with the longest possible baselines between sites. As the sensitive band of the noise curve is generally limited by available technology and fundamental noise sources, we are left with the option of dispersing the detectors as widely as possible.

Given the necessity to build detectors on Earth, the maximum possible baseline would be the diameter of the Earth, 42.5\,ms (distances converted to gravitational-wave travel times). The existing LIGO-Virgo network of detectors consists of three sites, at Hanford, Livingston and Cascina, all in the northern hemisphere. The longest baseline between detectors in this network, Hanford-Virgo, is 27\,ms. With the addition of the Gingin site to the network, this is increased to 41\,ms for the Livingston-Gingin baseline, close to the maximum possible, and the two baselines from Gingin to the other sites are all also of great length. The Japanese and Indian sites similarly increase the longest baseline, with Japan being closer to the U.S. sites but India closer to Virgo. All times are given for reference in table \ref{tab:time-delays}.

\begin{table}[htb]
\begin{tabular}{|r||c|c|c|c|c|c|} \hline
~ & H & L  & V  & A & I & J \\ \hline \hline
H & 0 & 10 & 27 & 39 & 36 & 25 \\ \hline
L & 10 & 0 & 26 & 41 & 39 & 32\\ \hline
V & 27 & 26 & 0 & 37 & 22 & 29 \\ \hline
A & 39 & 41 & 37 & 0 & 14 & 7\\ \hline
I & 36 & 39 & 22 & 14 & 0 & 21\\ \hline
J & 25 & 32 & 29 & 7 & 21 & 0 \\ \hline
\end{tabular}
\caption{\label{tab:time-delays}Table of gravitational-wave travel times between sites (Hanford, Livingston, Virgo, Australia, India, Japan, identified by their first letters), in milliseconds. The maximum possible baseline for a terrestrial network is 42.5\,ms.}
\end{table}

In this study, we compare the networks under the assumption that all detectors are operational at the time of the observation of a signal.  In practice, however, detectors have limited duty factors.  The probability of having at least three non-colocated detectors up at a given time is higher with the larger networks than with the HHLV network, increasing the probability for decent source localization.

Although the triangulation method above captures well the essential reason for desiring a longer baseline, the methods used in this paper are based on a fully coherent Bayesian analysis of the data to extract posterior probability functions on the parameters of interest. This method naturally incorporates the information from the time delays between sites, but it also includes the information from the amplitudes and relative phases of the signals present in each detector. This information can be used to further restrict the sky location, for example by eliminating the secondary maximum in the sky location for the majority of signals in the 3-detector network.

The addition to the network of a fourth site also gives a fourth separate detector orientation (instead of the replica of the H1 detector in the HHLV network). This raises the possibility of improved measurement of the other  parameters of the source, in particular, the measurement of the polarization angle $\psi$ and inclination angle $\iota$ of the gravitational wave may be expected to improve in certain cases. We perform a comprehensive comparison of measurement accuracies of these and other parameters encoded in the gravitational-wave signal from an inspiraling binary composed of non-spinning neutron stars.

\section{Analyses}

For this study, we used two independent Bayesian inference codes that implement two different techniques: the \LALInferenceMCMC code (based on the \SPINspiral code ~\cite{vanderSluys:2008a, vanderSluys:2008b, Raymond:2009}) uses a Markov Chain Monte Carlo algorithm (MCMC) ~\cite{Gilks:1996}, while the \Inspnest code~\cite{Veitch:2008wd,Veitch:2010} uses nested sampling \cite{Skilling:2006}.
%The first is designed to estimate the parameters of the signal assuming a gravitational wave is  present in the data. The second calculates the confidence in the presence of the signal, quantified by the {\it odds ratio} between the signal and noise models of the data.

Both techniques stochastically sample the parameter space in a search for the parameters that best match the observed data, simultaneously finding the set of parameters that yield the best fit to the data, and determining the accuracy of the parameter estimation.  This is achieved by calculating the posterior probability density function (PDF) on the parameter space $\vec{\theta}$ of the signal, given the data in the frequency domain $\tilde{d}$, and a signal model hypothesis $H$, which is
\begin{align}
p(\vec{\theta}|\tilde{d},H)&=\frac{p(\vec{\theta}|H)p(\tilde{d}|\vec{\theta},H)}{p(\tilde{d}|H)} \nonumber \\
&\propto p(\vec{\theta}|H) 
\exp \left(-\frac{1}{2} \langle \tilde{d}-\tilde{h}(\vec{\theta}) | \tilde{d}-\tilde{h}(\vec{\theta}) \rangle \right),
\label{eqn:posterior}\end{align}
where $p(\vec{\theta}|H)$ is the prior probability density of
the parameters $\vec{\theta}$ and $\tilde{h}(\vec{\theta})$ is the model used to
describe the signal in the frequency domain~\cite{Rover:2006ni}.  
The noise-weighted residuals in the presence of Gaussian noise with power spectral density
$S_\mathrm{n}(f)$ are given by 
\be
\langle \tilde{d}-\tilde{h}(\vec{\theta}) | \tilde{d}-\tilde{h}(\vec{\theta}) \rangle = 4 \int_0^\infty  \frac {\left|\tilde{d}(f)
    - \tilde{h}(f; \vec{\theta}) \right|^2}{S_\mathrm{n}(f)}\, df.
\ee
%\prod_k{}(\sqrt{2\pi}\sigma_k)^{-1}\exp\left(-\frac{\left[d_k-h_k(\vec{\theta})\right]^2}{2\sigma_k^2}\right)
For these simulations, for the LIGO and Virgo detectors, including the A and I sites, we used simulated noise with noise power spectral density $S_\mathrm{n}(f)$ similar to the Advanced LIGO design curve from the LIGO Algorithm Library (LAL) \cite{LAL}. Use of the Advanced LIGO noise curve for the Advanced Virgo detector may change the absolute results slightly, however the relative improvements ought to be consistent. The noise curve fit has a functional form of
\begin{eqnarray}
S_\mathrm{n}(f) = S_0\left\{  \left(\frac{f}{f_0}\right)^{-4.14} - 5\left(\frac{f_0}{f}\right)^2 \right. \nonumber \\ 
\left. + 111  \left(\frac{1 -\left(\frac{f}{f_0}\right)^2 + \frac{1}{2}\left(\frac{f}{f_0}\right)^4}{1 + \frac{1}{2}\left(\frac{f}{f_0}\right)^2} \right)\right\},
\label{eqn:AdLIGO}
\end{eqnarray}
where $S_0 = 10^{49}\,\mathrm{Hz}^{-1}$ and $f_0 = 215$\,Hz. For the LCGT detector, an interpolated LCGT design sensitivity curve was used~\cite{Kuroda:LCGT,Kuroda:LCGTnoisecurve}.

The MCMC method used by \LALInferenceMCMC explores the parameter space with a random walk, using the Metropolis-Hastings algorithm to simulate samples $\vec{\theta}$ from the posterior probability distribution function $p(\vec{\theta}|d,H)$. \LALInferenceMCMC uses a variety of optimisation techniques, including parallel tempering, to converge on the modes of the distribution and ensure adequate mixing of the chain~\cite{vanderSluys:2008a, vanderSluys:2008b, Raymond:2009}. It was started with randomly offset parameter values to simulate imperfect initial inputs from the detection pipeline, and 10 chains ran on each event to check for convergence.

The nested sampling algorithm used by \Inspnest operates by generating and replacing samples from the prior distribution $p(\vec{\theta})$, gradually shrinking the volume sampled by imposing a limit of minimum likelihood $p(d|\vec{\theta},H)$ on each replacement sample~\cite{Skilling:2006}. \Inspnest samples the prior distribution using an MCMC algorithm, which is optimised for exploring the structure of the limited prior, with jumps proposed through differential evolution, and along possible degeneracies in parameter space~\cite{Veitch:2008wd,Veitch:2010}. It is designed to compute the evidence $p(d|H)$, but the output samples from the prior can be weighted appropriately and resampled to produce samples from the posterior distribution.

The output from both Bayesian codes is a list of samples from the posterior distribution $p(\vec{\theta}|d,h)$, which are then used to estimate the mean, variance, and percentiles of the distributions. We performed two-dimensional binning and used a greedy algorithm to compute two-dimensional minimum probability intervals, as explained in section~\ref{sec:Simulations}.

As well as the two Bayesian methods, we also used the Fisher information matrix (FIM) to estimate the measurement uncertainties. The FIM technique, which approximates the likelihood surface quadratically near the likelihood peak, has a long history in gravitational-wave parameter estimation (e.g.,~\cite{CutlerFlanagan}).  It is known to suffer from a number of flaws, particularly in the low-SNR limit (when the quadratic approximation breaks down), or when correlations between parameters are very significant \cite{Vallisneri:2008}.  Furthermore, the Fisher information matrix is entirely local, and only approximates the shape of the maximum where it is evaluated, ignoring other maxima in the global parameter space which are picked up by the Bayesian analyses.  On the other hand, the FIM technique is computationally inexpensive, and hence allows a larger number of sources to be simulated in order to improve statistics.

The signal model for a non-spinning inspiral signal requires nine physical parameters: the chirp mass $\Mc=(m_1 m_2)^{3/5} (m_1+m_2)^{-1/5}$ and the symmetric mass ratio $\eta= m_1 m_2 (m_1+m_2)^{-2}$ (where $m_1$ and $m_2$ are the individual masses), right ascension $\alpha$, declination $\delta$, inclination $\iota$, orientation $\psi$, the luminosity distance $d_L$, the time of coalescence $t_c$ and the phase at coalescence $\phi$. For the Bayesian analyses, the prior probability distribution was assumed to be isotropic on the sphere of the sky, and on the inclination of the binary relative to the line of sight, and proportional to ${d_L}^2$. All other priors are uniform unless otherwise specified.

\section{Simulations} \label{sec:Simulations}

For this study, we assumed that a successful detection has already been made, and we have the correct waveform model to process the data, so we did not perform evidence calculations or model selection. The waveform model used for injection in all cases was generated by the LIGO Algorithm Library GeneratePPNInspiral routine~\cite{LAL}, which uses a time-domain approximant at 2.0 pN order in phase and 0 pN order in amplitude~\cite{Blanchet:1996}. All injections used symmetric binary neutron star signals with masses $m_1=m_2=1.4\,M_\odot$ as observed in the detector frame. For recovery and posterior calculation, we used the frequency-domain stationary phase approximation TaylorF2 approximant from LAL at 2.0 pN order (see, e.g., \cite{PNwaveforms:2009,Damour:2001} for additional information on post-Newtonian waveform approximants).

Injections were coherently made into four network configurations: (i) HHLV, (ii) AHLV, (iii) HILV, (iv) HHJLV. All injections were performed using Gaussian colored noise, using the Advanced LIGO power spectral density approximated by Eq.~(\ref{eqn:AdLIGO}) for the Advanced LIGO and Advanced Virgo sites (including Australia and India), and a fit to the LCGT noise curve for the J detector in the HHJLV network. In each of the four-site networks the noise realizations were kept the same in each network, although the sites are moved. In the HHJLV network the noise realisation in the HHLV portion was the same as in the other 4-site networks.  We also examined the possibility of rotating the Australia detector by $45^\circ$, and found very similar results to the AHLV case, which are omitted from further discussion for brevity.

The additional A and I detectors were positioned with their arms oriented along the local North and East vectors projected onto the tangent plane to the Earth's surface at the site, whose locations are given in section \ref{sec:Introduction}. The J detector was aligned with its y-arm $19^\circ$ from the local North vector as the geometry of the detector is constrained by the existing underground tunnels.

As the speeds of the methods differed, we performed a different set of simulations with each method, with the following details.

\subsubsection*{Nested Sampling}
The nested sampling implementation was able to run on a large set of injections. All signals were injected at a luminosity distance of 180 Mpc, but the inclination angle, polarization angle, right ascension and declination were located on a $5\times{}6\times{}5\times{}5$ rectangular grid in the $\alpha\times{}\sin\delta\times{}\cos\iota\times{}\psi$ parameter space for a total of 750 injections. This resulted in a range of SNRs between 5 and 35, but mostly between 7 and 25. We chose low and high frequency cutoffs of 30\,Hz and 2048\,Hz, which included the maximum frequency of the inspiral signal as no merger or ring-down components were used. Of the 750 signals, 728 were detected in all network configurations, and it is these we will use for the summary statistics throughout the remainder of the paper.
The nested sampling search, by its nature, samples the entire prior range given for the parameters. In this case, the total mass $M=m_1+m_2$ was assumed to be between 2 and 35 $\Msun$, with component masses allowed to be in the range $m_1,m_2 \in [1,30]\,\Msun$. The prior distribution on $\Mc$ used was $p(\Mc)\propto{}d_L^2\Mc^{-5/6}$, chosen as an approximation to the Jeffrey's prior which sets the prior as a function of the Fisher matrix $\Gamma$ on the parameter space, $p(\vec{\theta})\propto\sqrt{\det\Gamma(\vec{\theta})}$, in order to improve sampling where the template bank density is highest \cite{Jeffreys}.

\subsubsection*{MCMC}
The MCMC method was used to run on a randomly chosen subset of the injections that were analyzed with nested sampling (computational constraints prevented us from using the full set of injections).  Results from $42$ injections are used in this analysis.  For consistency, the MCMC runs employed the same priors and frequency range for the overlap integral as the nested sampling runs.  The MCMC results on individual injections matched the nested sampling results, therefore allowing us to gain extra confidence in the Bayesian parameter estimation.  However, because of the concern that smaller numbers of runs could increase statistical fluctuations, we do not quote absolute accuracies for MCMC runs, but only compare the expected parameter estimation accuracies for different network configurations.

\subsubsection*{Fisher Matrix}

The FIM has the advantage of being computationally inexpensive, and so permits a large Monte Carlo over injections. Therefore, we used the FIM to confirm the results of our Bayesian analyses.    We used a low-frequency cutoff of $30$ Hz and integrated up to the innermost stable circular orbit frequency, around $1600$ Hz.  

For this study, we varied all angles in a Monte Carlo of 4000 points, adjusting the distance to keep the total network SNR equal to $30$ for all injections in all network configurations.   We chose injections with this relatively high value of SNR, rather than the SNR distribution used for nested sampling and MCMC studies, because of concerns about the the accuracy of the FIM approach outside the high-SNR limit.  The SNRs were separately normalized to $30$ for all networks, including the five-detector HHJLV network; therefore, results for that network from the FIM study are expected to be worse than those  obtained with Bayesian studies, which would normally have higher SNRs in the five-detector network than the four-detector networks for a given injection.

\subsection{Quantities compared}

Although our techniques make full nine-dimensional posterior PDFs available, these are unwieldy to compare or visualize. So we typically consider only one- or two-dimensional PDFs marginalized over the remaining parameters, with examples shown in figures \ref{fig:degeneracy} \& \ref{fig:PDFcomp}. However, to allow us to make comparisons, we had to restrict ourselves to particular estimators for the PDFs, with the understanding that unless PDFs are extremely narrow or are described by a simple analytical function (e.g., a Gaussian), a few estimators are not sufficient to describe all of the information contained in the PDFs.

We estimated the width of a particular one- or two-dimensional PDF as follows.  For a given fraction $0<F<1$, the $F$-width of a one-dimensional PDF was defined as the width of the smallest region that contains that fraction $F$ of the posterior PDF. 
%Thus, the $1\sigma$, $2\sigma$, and $3\sigma$ widths represent the width of the smallest regions that contain $68.3\%$, $95.4\%$, and $99.73\%$ of the total posterior probability, respectively.
Thus, the 95\% width represents the width of the smallest region of parameter space that contains 95\% of the total posterior probability. A similar approach was used for two-dimensional PDFs in (inclination, distance) space and (right ascension, declination) space, with pixels of a fixed size ($0.25$ deg$^2$) being used in a greedy algorithm to estimate the sky area for sky localization.  

We also define a ``standard accuracy'' (by analogy with the standard deviation) as the square root of the mean of the sum of the squared differences between the points in the PDF (sampled according to the posterior) and the true value.  Thus, for a marginalized one-dimensional PDF, 
$$standard\ accuracy = \sqrt{\frac{1}{N}\sum_{i=1}^N (x_i-x_{\rm true})^2}.$$
For a PDF whose mean is equal to the true value, the standard accuracy is just the standard deviation.  For a delta-function PDF that is biased away from the true value, the standard accuracy is the error.  In general, the standard accuracy is equal to the sum, in quadrature, of the standard deviation of the PDF and the difference between the PDF mean and the true value.

\subsection{Relative Improvements}

% Results, comparisons between networks
In the following sections, we present the results of each of the three analyses of the differences in parameter estimation accuracy between the three networks. In the case of the two Bayesian codes, we measure the 95\% confidence intervals for each parameter and compute the ratio of these in comparison to the result from the HHLV network for the same injection. The median value of the ratio is quoted in each table, along with the range incorporating the 5\% and 95\% quantiles of the distribution. This gives some measure of the spread of the ratios across different sky positions, locations and orientations.

%%%%%%%%%%%%%%% RESULTS TABLES %%%%%%%%%%%%%%%%%%%%%%%%%
\begin{table*}[h!tb]   
\begin{tabular}{c@{\quad\vline\quad}c@{\quad}c@{\quad\vline\quad}c@{\quad}c@{\quad\vline\quad}c@{\quad}c}
\hline
& \multicolumn{2}{c}{AHLV / HHLV} &  \multicolumn{2}{c}{HILV / HHLV}  &  \multicolumn{2}{c}{HHJLV / HHLV}\\
Parameter  & $95\%$ width  & $std.~acc$ & $95\%$ width & $std.~acc $ & $95\%$ width & $std.~acc $\\\hline

$\Mc$ & $ 0.97^{+2.03}_{-0.66} $ & $ 0.93^{+2.59}_{-0.66}$ 
& $ 1.00^{+2.18}_{-0.65} $ & $ 0.85^{+2.27}_{-0.73}$ 
& $ 0.82^{+1.92}_{-0.80} $ & $ 0.72^{+3.23}_{-0.69}$ 
\\
$\eta$ & $ 0.93^{+3.27}_{-0.70} $ & $ 0.94^{+3.34}_{-0.80}$ 
& $ 0.96^{+1.26}_{-0.59} $ & $ 0.80^{+1.63}_{-0.55}$ 
& $ 0.88^{+1.72}_{-0.63} $ & $ 0.77^{+3.61}_{-0.50}$ 
\\
$t_c$ & $ 0.62^{+1.06}_{-0.47} $ & $ 0.46^{+1.02}_{-0.41}$ 
& $ 0.71^{+1.86}_{-0.51} $ & $ 0.62^{+1.22}_{-0.56}$ 
& $ 0.55^{+0.60}_{-0.42} $ & $ 0.37^{+1.98}_{-0.32}$ 
\\
$d_L$ & $ 0.93^{+0.29}_{-0.23} $ & $ 0.98^{+0.11}_{-0.36}$ 
& $ 0.93^{+0.38}_{-0.24} $ & $ 0.96^{+0.17}_{-0.35}$ 
& $ 0.85^{+0.40}_{-0.27} $ & $ 0.95^{+0.19}_{-0.33}$ 
\\
$\alpha$ & $ 0.50^{+1.23}_{-0.29} $ & $ 0.43^{+0.77}_{-0.41}$ 
& $ 0.59^{+0.53}_{-0.46} $ & $ 0.47^{+1.87}_{-0.45}$ 
& $ 0.50^{+0.83}_{-0.43} $ & $ 0.46^{+1.47}_{-0.44}$ 
\\
$\delta$ & $ 0.43^{+0.74}_{-0.35} $ & $ 0.27^{+0.98}_{-0.23}$ 
& $ 0.50^{+0.70}_{-0.38} $ & $ 0.46^{+0.97}_{-0.43}$ 
& $ 0.55^{+0.62}_{-0.42} $ & $ 0.29^{+1.67}_{-0.24}$ 
\\
$\iota$ & $ 0.85^{+0.51}_{-0.31} $ & $ 1.01^{+0.64}_{-0.56}$ 
& $ 0.88^{+0.42}_{-0.52} $ & $ 1.00^{+0.35}_{-0.41}$ 
& $ 0.82^{+0.68}_{-0.45} $ & $ 1.04^{+0.54}_{-0.47}$ 
\\
$\psi$ & $ 0.98^{+0.38}_{-0.60} $ & $ 0.98^{+0.16}_{-0.17}$ 
& $ 0.97^{+0.38}_{-0.66} $ & $ 0.99^{+0.11}_{-0.25}$ 
& $ 0.95^{+0.55}_{-0.63} $ & $ 0.99^{+0.17}_{-0.15}$ 
\\
$\alpha-\delta$ & $ 0.27^{+0.65}_{-0.21} $ & ---
& $ 0.30^{+0.89}_{-0.22} $ & ---
& $ 0.38^{+0.77}_{-0.29} $ & ---
\\
$d_L-\iota$ & $ 0.76^{+0.49}_{-0.32} $ & ---
& $ 0.80^{+0.36}_{-0.54} $ & ---
& $ 0.76^{+0.57}_{-0.51} $ & ---
\\
\hline
\end{tabular}
\caption{Comparative $95\%$ interval widths and standard accuracies for one-dimensional PDFs, and comparative $95\%$ areas for two-dimensional PDFs (last two lines) averaged over all injections, calculated using the MCMC algorithm.  All values are reported as fractions of the corresponding values for the HHLV network configuration.  The median values of the ratios across all injections are computed; the error bars correspond to the spread between the 5\% and 95\% quantile values of these ratios across all injections. See text for details.   
%All 9 runs are included. 
\label{mean9}}

\begin{tabular}{c@{\quad\vline\quad}c@{\quad}c@{\quad\vline\quad}c@{\quad}c@{\quad\vline\quad}c@{\quad}c}
\hline
& \multicolumn{2}{c}{AHLV / HHLV} &  \multicolumn{2}{c}{HILV / HHLV}  &  \multicolumn{2}{c}{HHJLV / HHLV}\\
Parameter  & $95\%$ width  & $std.~acc$ & $95\%$ width & $std.~acc $ & $95\%$ width & $std.~acc $\\
\hline
$\Mc$ & $ 1.00^{+0.80}_{-0.40} $ & $ 1.00^{+1.92}_{-0.56}$ 
& $ 1.00^{+0.71}_{-0.47} $ & $ 1.02^{+1.39}_{-0.68}$ 
& $ 0.92^{+0.33}_{-0.37} $ & $ 0.98^{+0.34}_{-0.42}$ 
\\
$\eta$ & $ 1.00^{+0.78}_{-0.38} $ & $ 1.02^{+1.28}_{-0.51}$ 
& $ 1.00^{+0.70}_{-0.44} $ & $ 1.01^{+1.14}_{-0.56}$ 
& $ 0.92^{+0.30}_{-0.29} $ & $ 0.98^{+0.35}_{-0.38}$ 
\\
$t_c$ & $ 0.73^{+0.54}_{-0.47} $ & $ 0.69^{+1.04}_{-0.61}$ 
& $ 0.69^{+0.61}_{-0.46} $ & $ 0.62^{+0.90}_{-0.52}$ 
& $ 0.68^{+0.32}_{-0.43} $ & $ 0.66^{+0.49}_{-0.57}$ 
\\
$d_L$ & $ 1.00^{+0.33}_{-0.21} $ & $ 0.98^{+0.15}_{-0.25}$ 
& $ 1.05^{+0.45}_{-0.30} $ & $ 0.91^{+0.53}_{-0.30}$ 
& $ 0.92^{+0.25}_{-0.24} $ & $ 0.98^{+0.16}_{-0.30}$ 
\\
$\alpha$ & $ 0.67^{+0.58}_{-0.48} $ & $ 0.61^{+0.88}_{-0.55}$ 
& $ 0.62^{+0.52}_{-0.47} $ & $ 0.56^{+1.00}_{-0.50}$ 
& $ 0.60^{+0.40}_{-0.44} $ & $ 0.56^{+0.59}_{-0.52}$ 
\\
$\delta$ & $ 0.50^{+0.70}_{-0.40} $ & $ 0.39^{+1.12}_{-0.34}$ 
& $ 0.61^{+0.59}_{-0.48} $ & $ 0.48^{+1.02}_{-0.41}$ 
& $ 0.67^{+0.33}_{-0.49} $ & $ 0.59^{+0.53}_{-0.49}$ 
\\
$\iota$ & $ 0.93^{+0.24}_{-0.46} $ & $ 0.91^{+0.35}_{-0.58}$ 
& $ 0.86^{+0.28}_{-0.47} $ & $ 0.83^{+0.56}_{-0.56}$ 
& $ 0.86^{+0.19}_{-0.44} $ & $ 0.90^{+0.26}_{-0.59}$ 
\\
$\psi$ & $ 1.00^{+0.19}_{-0.52} $ & $ 0.99^{+0.32}_{-0.51}$ 
& $ 0.97^{+0.11}_{-0.66} $ & $ 0.91^{+0.32}_{-0.63}$ 
& $ 0.93^{+0.11}_{-0.52} $ & $ 0.97^{+0.22}_{-0.56}$ 
\\
$\alpha - \delta$ & $ 0.32^{+0.78}_{-0.24} $ & ---
& $ 0.43^{+0.87}_{-0.32} $ & ---
& $ 0.33^{+0.51}_{-0.22} $ & ---
\\
$d_L - \iota$ & $ 0.90^{+0.29}_{-0.45} $ & ---
& $ 0.88^{+0.44}_{-0.49} $ & ---
& $ 0.53^{+0.24}_{-0.27} $ & ---
\\\end{tabular}

\caption{\label{tab:inspnest}Comparative 95\% probability interval widths and standard accuracies between alternative network configurations and the HHLV network, calculated using the nested sampling algorithm. The median values of the ratios across 728 injections detected in all networks are computed. Errors quoted correspond to the 5\% and 95\% quantiles of the distribution of ratios, as in table \ref{mean9}. For $\alpha-\delta$ plane, 95\% probability intervals are calculated using a greedy binning algorithm.}

\begin{tabular}{c@{\quad\vline\quad}c@{\quad\vline\quad}c@{\quad\vline\quad}c}
\hline
Parameter & AHLV/HHLV & HILV/HHLV & HHJLV/HHLV\\
\hline
$\Mc$ &  1.00 & 1.00 & 1.02 \\
$\eta$ &  1.00 & 1.00 & 1.02\\ 
$t_c$ (msec) &  0.54 & 0.56 & 0.65\\    
$d_L$      &      0.77 & 0.66 & 0.71\\
$\alpha$ (deg)    &      0.66    &  0.56 & 0.50\\
$\delta$ (deg)   &   0.31 & 0.57 & 0.62\\
$\iota$ (deg)   &  0.76 & 0.67 & 0.70\\   
$\psi$ (deg)   &     0.77    &  0.67 & 0.69 \\
$\phi_c$ (deg) &    0.88 & 0.84 & 0.91\\
$\alpha-\delta$ (deg$^2$) & 0.29 & 0.35 & 0.47\\

\hline
\end{tabular}
\caption{\label{tab:fisher}Ratios of median standard deviations for each parameter and the sky area as reported by the Fisher information matrix, as a function of network configuration.  All injections are re-normalized to an SNR of 30 in the given network, so the last column, corresponding to a five-detector network re-normalized to the same SNR as the four-detector networks, can not be directly compared to similar columns in the tables above.}

\end{table*}

\subsubsection*{MCMC results}

In table \ref{mean9}, we average comparisons of the $95\%$ confidence intervals across the 42 injections analyzed with the \LALInferenceMCMC code.   We show the values of the $95\%$ confidence interval widths for the extended network configurations as fractions of the same widths for the HHLV configuration.  Table \ref{mean9} lists the median ratio of the size of the $95\%$ probability region over all of our injection runs, while the $5\%$ and $95\%$ percentiles of the distribution of ratios are shown in super and subscript, to indicate the range of the results.
% We list the mean values, and also include minimum and maximum interval ratios to indicate the spread due to different sky locations, inclinations, and orientations, as well as different noise realizations.  

As discussed in more detail below, the accuracy with which individual masses can be recovered is not significantly affected by the network configuration.  This conforms to our expectation that mass measurements, which come from waveform phasing, are constrained primarily by the total network SNR. Sky localization accuracy can be significantly improved in both directions when a fourth site is added to the network.  Timing accuracy at the geocenter is strongly correlated with sky localization and is similarly improved.  A fourth site also moderately improves the accuracy of distance measurements. 

We should particularly point out the next-to-last line of the table, ``$\alpha-\delta$''.  The area of this 2-dimensional PDF is a direct measure of the uncertainty in estimating the position of the source on the sky.  The error box shrinks by a factor of $\sim 4$ when the second Hanford detector is moved to Australia or India because of the much-improved north-south baseline.  This improvement in sky localization accuracy will make the detection of an electromagnetic counterpart to the gravitational-wave source more feasible, and is perhaps the biggest boon in moving one of the Hanford detectors to India or Australia.  

%\begin{figure}[htb]
%\centering
%\includegraphics[keepaspectratio=true, width=\columnwidth]{figures/HHLV-RA-dec.pdf}
%\includegraphics[keepaspectratio=true, width=\columnwidth]
%{figures/HLVA-RA-dec.pdf}
%\caption{Sky localization of a typical source with the HHLV (top) and AHLV (bottom) network configurations.  The $1\sigma$, $2\sigma$, and $3\sigma$ confidence areas are shown in red, yellow, and blue, respectively, while the true location is indicated with the target symbol.  Note the two allowed areas on the sky in the top plot vs.~a single area in the bottom plot, indicating a degeneracy that exists in a network with three sites but is removed when an interferometer is moved to Australia.}
%\label{fig:sky}
%\end{figure}

\subsubsection*{Inspnest results}

\begin{figure}
\resizebox{\columnwidth}{!}{\includegraphics{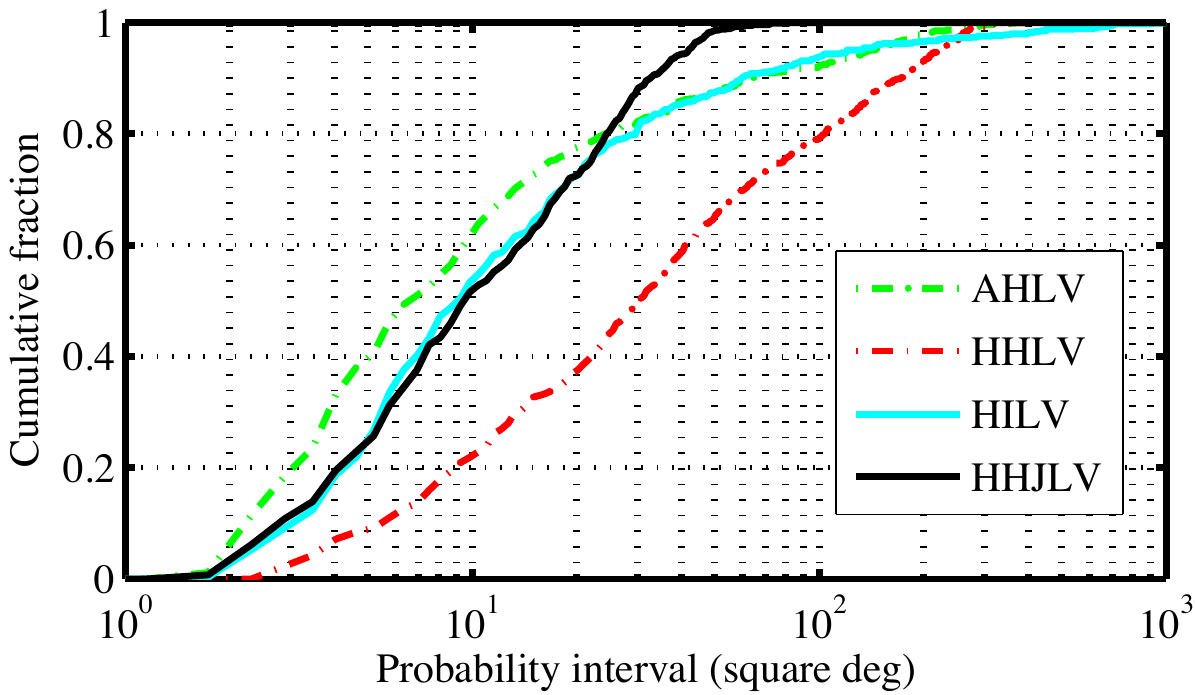}}
\caption{\label{fig:inspnest90pc}Cumulative histogram of the 95\% confidence interval for the area of the sky in square degrees, estimated using the \Inspnest analysis of 728 signals which were detected in all network configurations, covering the range of sky locations and orientations of the binary. Different lines correspond to different networks: green dot-dashed line, AHLV; red dashed line, HHLV; cyan solid line, HILV; black solid line HHJLV. The extended networks show significantly better performance than the 3-site network, with AHLV offering the highest fraction of signals resolved to better than 10 square degrees. HILV and HHJLV show similar performance to each other for a large fraction of the signals, but the HHJLV network avoids the tail of poorly resolved signals located in regions of parameter space with poor sensitivity in HILV.}
\end{figure}

%\begin{figure}
%\resizebox{\columnwidth}{!}{\includegraphics{figures/HHLV_probability_hist.pdf}}
%\caption{\label{fig:inspnest-hist}The distribution of \Inspnest quantiles of injected values of the right ascension within the posterior PDFs for this parameter, as estimated for the AHLV network.  This distribution should be uniform in order for the frequentist and Bayesian interpretations of confidence intervals to agree.}
%\end{figure}

In figure \ref{fig:inspnest90pc} we show the results of determining the sky location of 728 injections with the \Inspnest nested sampling code. The lines show the cumulative histograms for the 95\% confidence interval on the sky location of the binary for each of the network configurations considered.
As expected, the addition of a fourth site to the network yields a significant improvement in the resolution of the sky location, due to the increased baseline for relative timing measurements. The greatest improvement is seen for the addition of the the Australian detector, yielding the longest north-south baseline. HILV and HHJLV are both great improvements on the HHLV network, and have similar performance for the majority of signals, although HHJLV suffers from fewer outliers with sky resolution greater than 100 square degrees, possibly due to the increased signal-to-noise ratio from the additional detector.

This improved performance is reflected in the comparison of the median 95\% confidence interval for each of the networks, given in table \ref{tab:inspnest_medians}. The median resolution for all sources shrinks from 30.25 square degrees to 6.625 square degrees between HHLV and the most accurate AHLV network.   In addition to the reduction of the median resolution, these improvements are also reflected in the relative resolutions of the networks. Table \ref{tab:inspnest} shows the median \emph{ratio} of the 95\% probability region size for the extended networks against HHLV; note that the median of the ratios for the sky location intervals ($\alpha-\delta$ row) is not the same as the ratio of the medians in table \ref{tab:inspnest_medians}. 

The qualitative behavior inferred from the \LALInferenceMCMC and \Inspnest runs is in excellent agreement.  Some differences between the ``typical'' results tables \ref{tab:inspnest} and \ref{mean9} are due to the impact of statistical fluctuations in the smaller subset of injections analyzed with \LALInferenceMCMC; however, we have confirmed quantitative agreement between the two methods on the individual injections being analyzed.

%The sky localization improvements appear to be somewhat more modest in the \Inspnest results than in the \LALInferenceMCMC ones.  However, a quantitative analysis is hampered by the small number of \LALInferenceMCMC runs considered, and the differences between run configurations, while the qualitative behaviour is in excellent agreement.

For a more detailed look at the performance of the networks for the non-sky-location parameters, we show in figure \ref{fig:NS_CDF} the cumulative probability distributions for the fraction of signals found within a given probability interval width for $\Mc$, $\eta$, $d_L$, $\iota$, $\psi$ and $t_c$.
Although our range of injections at constant distance is astrophysically unrealistic, these figures give a good idea of the relative performance of the networks across an isotropic distribution of sky position, polarisation and inclination angles. We can see immediately that the improvement in the resolution of the chirp mass and $\eta$ parameters is marginal, as evident from tables \ref{tab:inspnest}. The slightly improved performance for HHJLV can be attributed to the higher signal-to-noise ratio in that network thanks to the fifth detector.

With the distance and inclination angle parameters, we see slight improvements, with the biggest effect again being produced by the additional SNR in the HHJLV network. Due to degeneracies in the parameter space, the effect is more pronounced when the two dimensional distribution is considered, as in section \ref{ss:degeneracy}.
From table \ref{tab:inspnest}, we see that the median relative improvement of the polarization angle $\psi$ resolution is minimal, however in figure \ref{fig:NS_CDF} it is apparent that for the 4-site networks there are a greater fraction of injections which are resolved well. This is explained by the large fraction of signals ($\sim 50\%$) which have a very broad distribution where the $\psi$ parameter is degenerate with the phase of the signal, when $\iota$ is close to 0 or $\pi$. Although the median of the improvement ratio for individual sources is $\sim 1$, the mean is $\sim 0.95$ for HALV and $\sim~0.85$ for HILV and HHJLV, indicating that the sources which can be resolved in $\psi$ are better resolved in the 4-site networks.
The $t_c$ distribution shows similar performance for all 4-site networks for most signals, with slightly fewer signals poorly resolved in time with the HILV network than AHLV, reflecting the shorter tails of the distributions in figures \ref{fig:inspnest90pc} and \ref{fig:Fisher-sa} (sky localization is strongly correlated with the accuracy of timing the gravitational wave at the geocenter).

\begin{table}[h!]
\begin{tabular}{|l|c|}
\hline
Network & Median 95\% conf. int. \\
\hline
AHLV & 6.625 deg$^2$ \\
HHLV & 30.25 deg$^2$ \\
HILV & 9 deg$^2$ \\
HHJLV & 9.5 deg$^2$ \\
\hline
\end{tabular}
\caption{\label{tab:inspnest_medians}Median 95\% confidence intervals in square degrees for each network configuration.}
\end{table}

\subsubsection*{Fisher matrix results}

\begin{figure}[htb!]
\centering
\includegraphics[keepaspectratio=true, width=\columnwidth, angle=0]{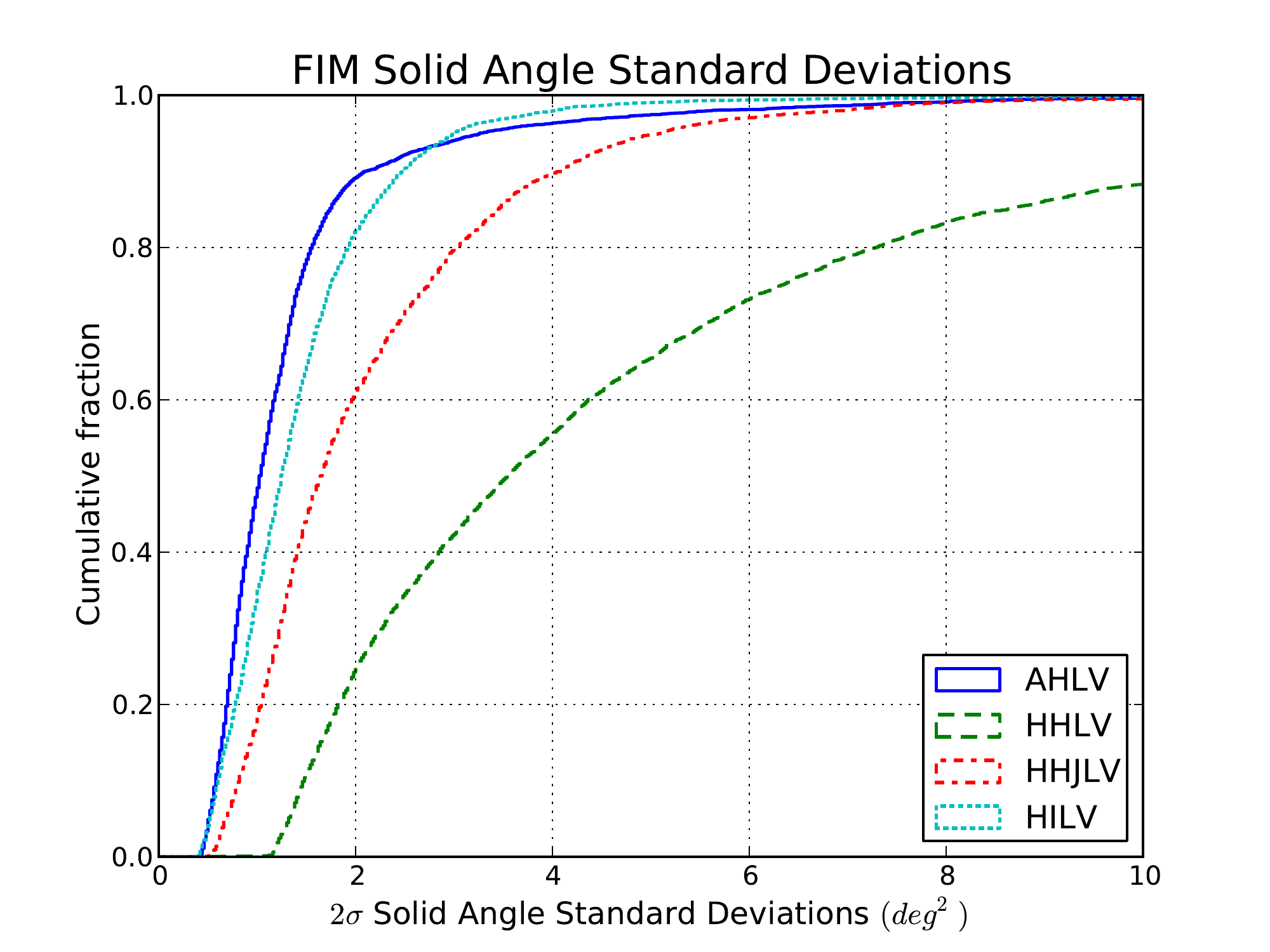}%{figures/Fisher-sa}
\caption{A cumulative histogram over injections of the estimated accuracies in sky localization obtained via the Fisher information matrix technique.  Blue, green, red and cyan are the distributions of sky-localization uncertainties for the AHLV, HHLV, HHJLV and HILV networks, respectively.  All injections are independently normalized to an SNR of 30 for each network configuration, to avoid concerns about the trustworthiness of FIM results at low SNRs.  The relative shapes of the histograms are thus more relevant than their actual values.}
\label{fig:Fisher-sa}
\end{figure}

The Fisher information matrix results are presented in table \ref{tab:fisher}.  These are ratios of the median standard deviations for each parameter, taken from a Monte Carlo over angles (with distance adjusted to keep the total network SNR equal to 30).  According to the FIM results, the greatest change in sky localization comes from the improvement in the measurement of the declination angle with a network that includes an Australian detector.  The declination angle can be measured more accurately by a factor of $\sim 3$ with an AHLV network relative to the HHLV network, as a detector in the Southern hemisphere greatly improves the latitudinal baseline of the network, allowing for superior angular resolution in that direction.  The last line in table \ref{tab:fisher} indicates the Fisher-matrix estimate of the sky area, defined following Eq.~(43) of \cite{BarackCutler}.  Figure \ref{fig:Fisher-sa} shows a histogram of the sky-area accuracy for two network configurations.

%The Fisher information matrix is insensitive to the antipodal degeneracy in sky location, because that degeneracy is global, while the FIM local.  So the FIM only describes the size of one of the two islands on the sky to which a source can be localized with three detectors.  As we have seen from the Bayesian results, adding a detector in Australia to the network yields a factor of $\sim 2$ improvement just by removing one of the two islands.  Thus, the FIM understates the improvement in sky location by approximately the same factor; once this is taken into account, the FIM results for sky localization agree quite well with the Bayesian results, prediction an improvement of approximately a factor of {\bf CHECK $4.5$} between the HHLV and AHLV networks.

% Comment on the results in the previous sections

\subsection{Parameter Degeneracies}\label{ss:degeneracy}

% Taken from the sections above

\begin{figure*}
\centering
\includegraphics[width=\columnwidth,height=8cm]{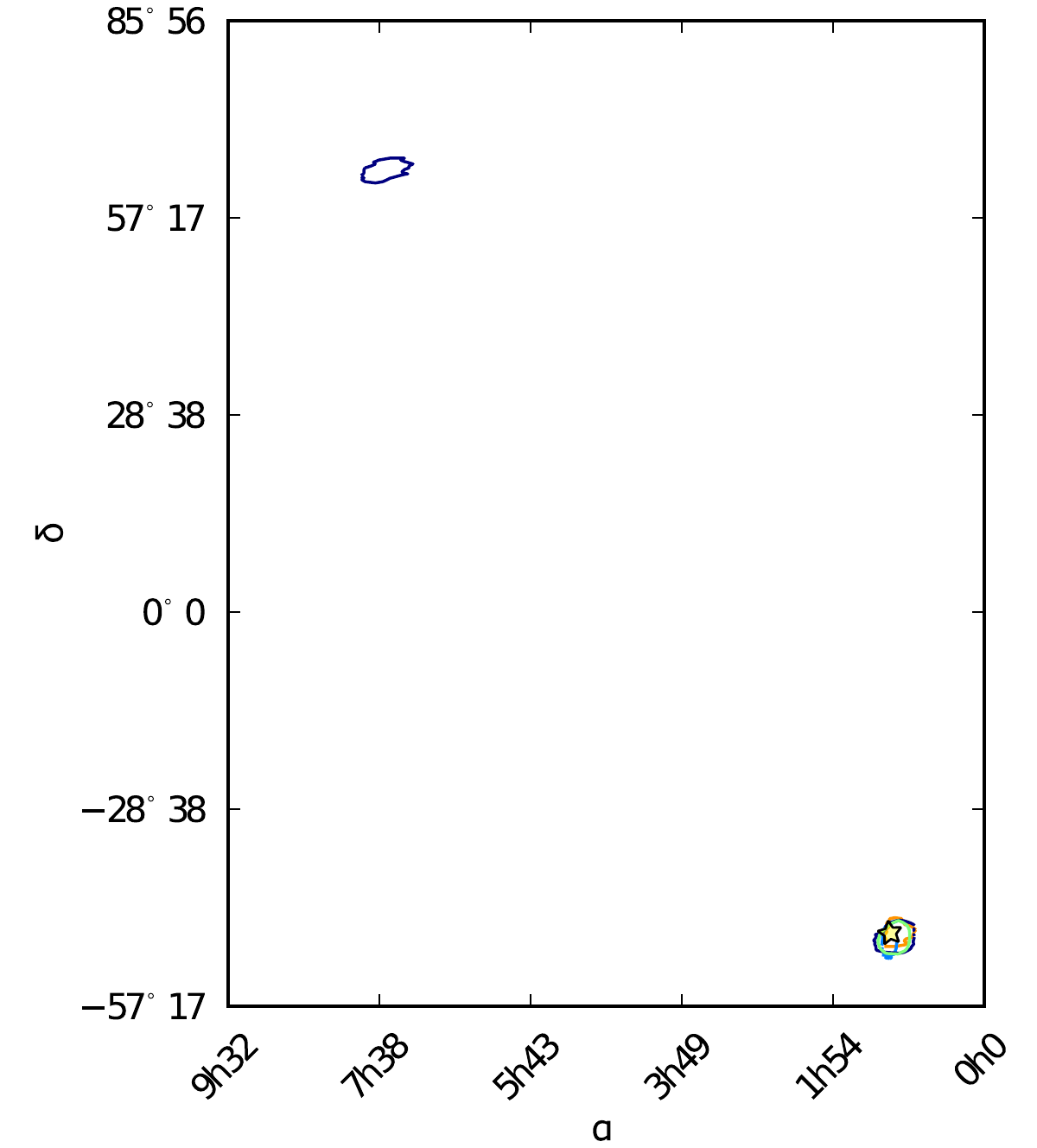}
\includegraphics[width=\columnwidth,height=8cm]{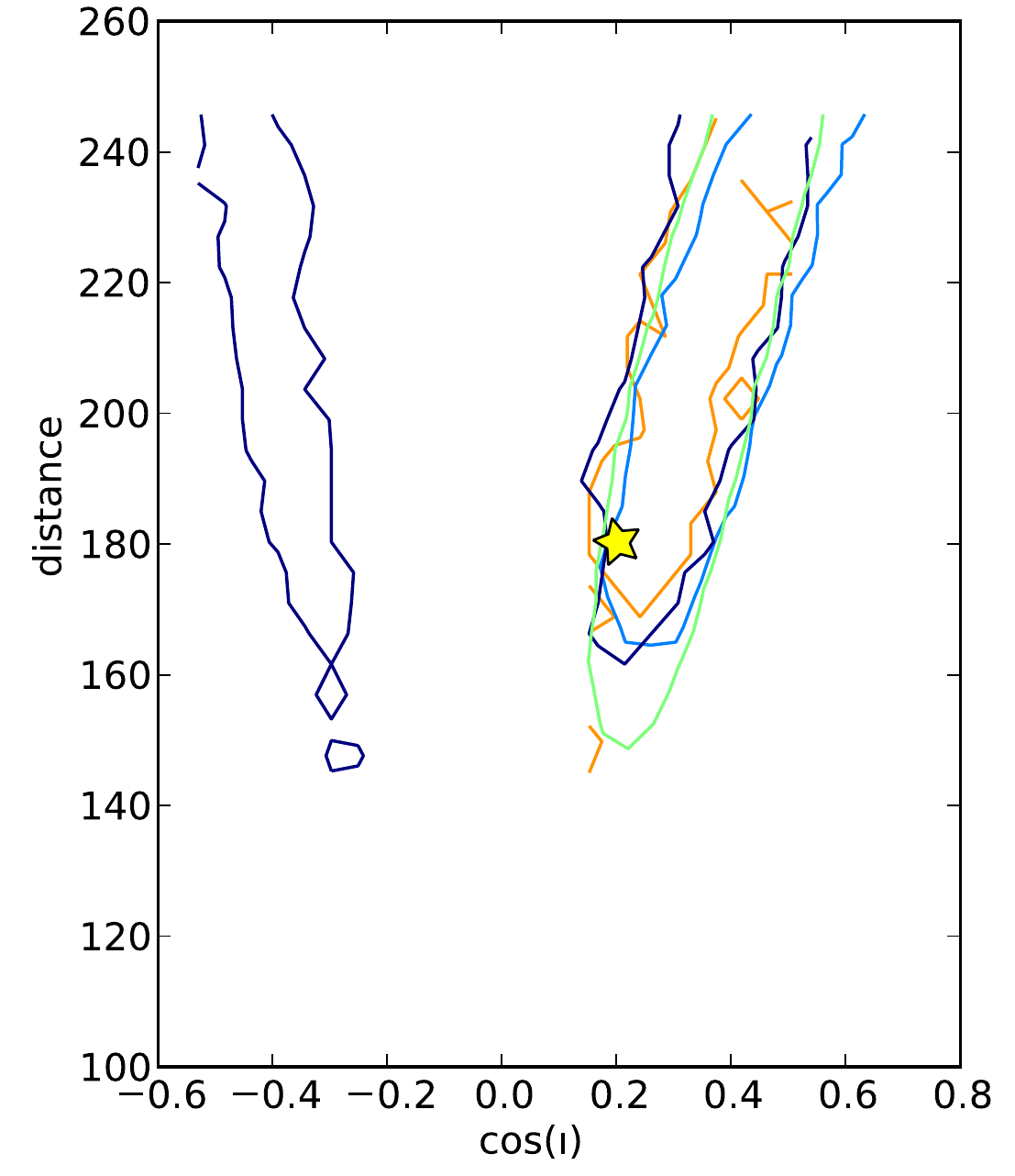}
\caption{\label{fig:degeneracy}An example of the ability of a four-site network to break the degeneracy between multiple modes in $\alpha-\delta$ and $\cos\iota-d_L$ parameter space. This example shows the 95\% confidence intervals from the HHLV (purple), AHLV (blue), HILV (green) and HHJLV (orange) networks. The sky position can be confined to one region with the four-site network, while the partial inclination angle degeneracy upon reflection is broken. The location of the injection in parameter space is indicated with a star.} 
\end{figure*}

In cases where a signal is observed with marginal SNR in one or more detectors, or when the orientation of the binary is unfavourable (i.e. $\iota$ is close to $90^\circ$), a degeneracy between two sky locations often emerges. The addition of the fourth site to the network ensures that the source can be localised to a single region on the sky, as shown in the 2-dimensional PDFs of figure \ref{fig:degeneracy} which compares the 95\% probability regions for different networks but the same injection.
Note the two areas on the sky in the left panel, indicating a degeneracy that exists in a network with three sites but is removed when a fourth site is added.
%The two-fold improvement in the time of coalescence, as defined by the time of arrival of the gravitational wave at the center of the Earth, is a reflection of this degeneracy breaking.

\begin{figure}[htb!]
\centering
\includegraphics[height=\columnwidth, angle=270]{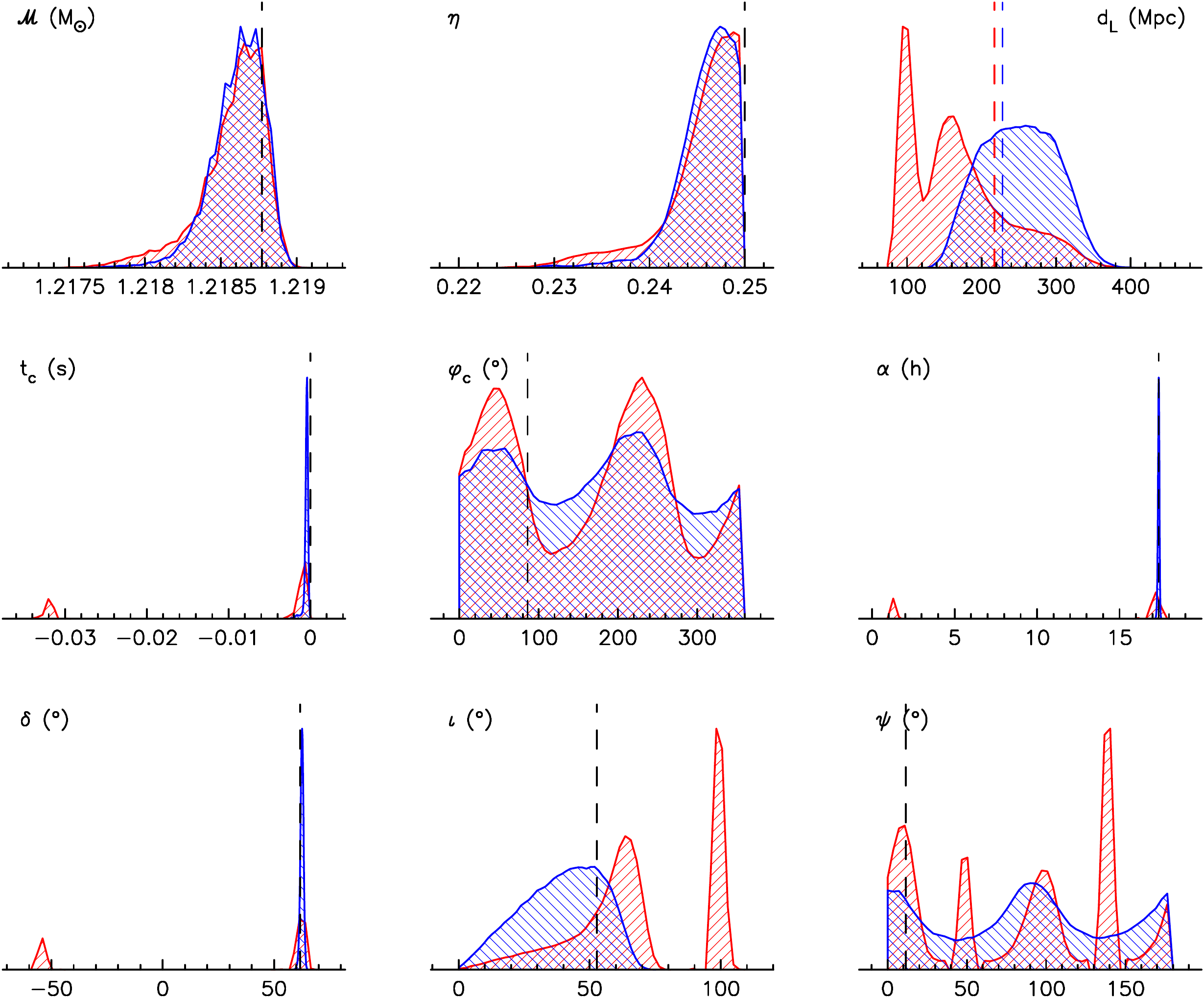}
\caption{Comparison of the one-dimensional PDFs for a typical source as detected by the HHLV network (red) and AHLV network (blue).  Note the bimodal posteriors in right ascension and declination for the HHLV network vs.~unimodal ones for the AHLV network.  The latter network also allows for better estimates of the posteriors for inclination and luminosity distance, which is not properly reflected by the simple estimators of the PDF width used in table \ref{mean9}.  Dashed lines indicate the true injected values (different true values of the luminosity distance were used for the HHLV and AHLV injections so that the total network SNR is 15 in both cases).
}
\label{fig:PDFcomp}
\end{figure}

Moderate improvements also appear in inclination and luminosity distance measurements.  Here, the one-dimensional estimators used in Tables \ref{mean9}, \ref{tab:inspnest} and \ref{tab:fisher} do not tell the full story because of the strong correlation between inclination and luminosity distance. The secondary maximum on the sky position also corresponds to a secondary mode in the $\cos\iota$ parameter corresponding to the transformation $\iota\mapsto\pi-\iota$.
In fact, the addition of a fourth site allows this degeneracy to be broken, as can be seen by comparing the marginalized one-dimensional PDFs for a sample injection the HHLV and AHLV network configurations, plotted in figure~\ref{fig:PDFcomp}.  Although the width did not shrink significantly from red to blue posteriors, with the AHLV network configuration the posteriors are unimodal and centered on the true values, as a degeneracy in the inclination-distance space is broken.

\begin{figure*}[htb!]
\centering
\includegraphics[width=\columnwidth]{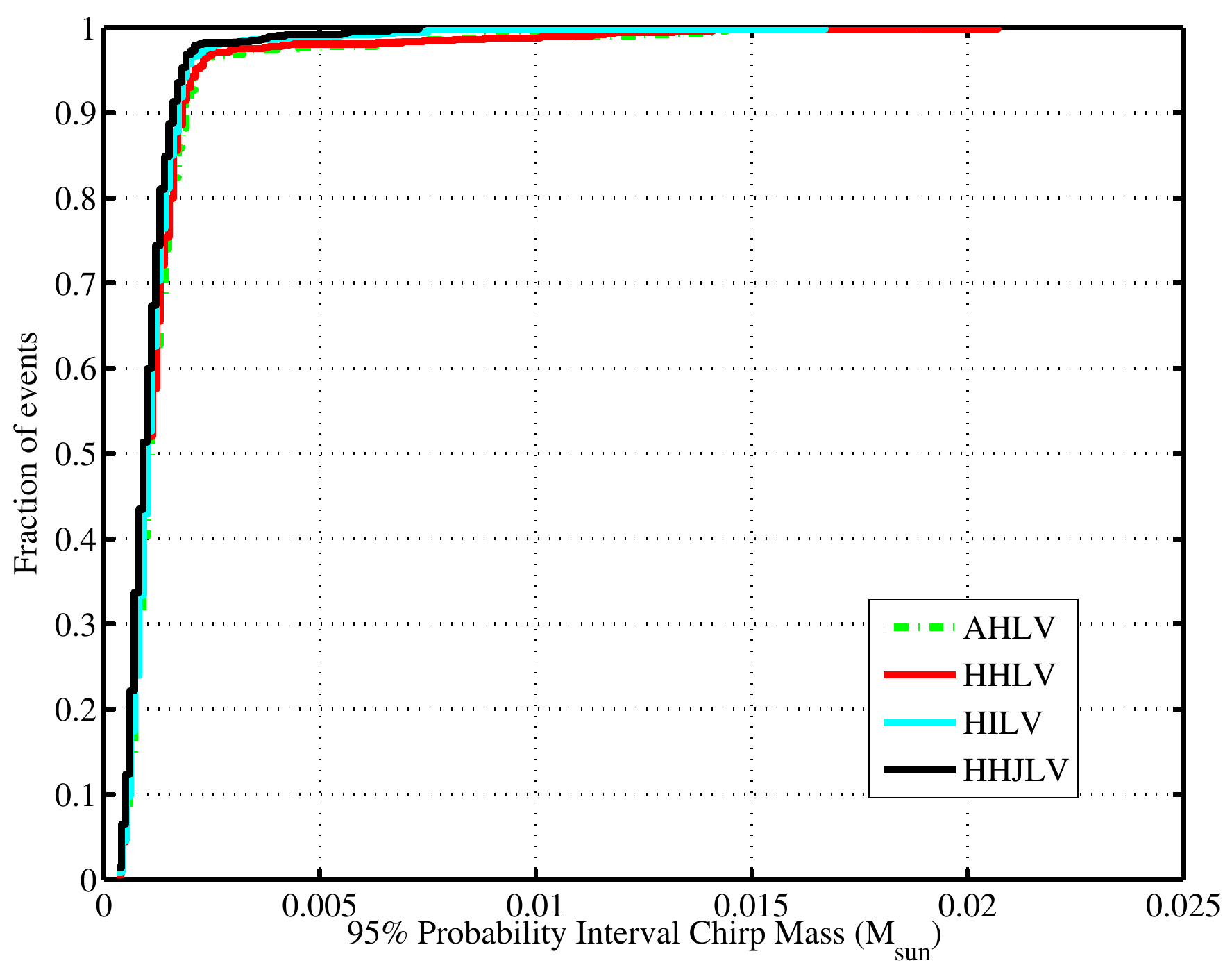}
\includegraphics[width=\columnwidth]{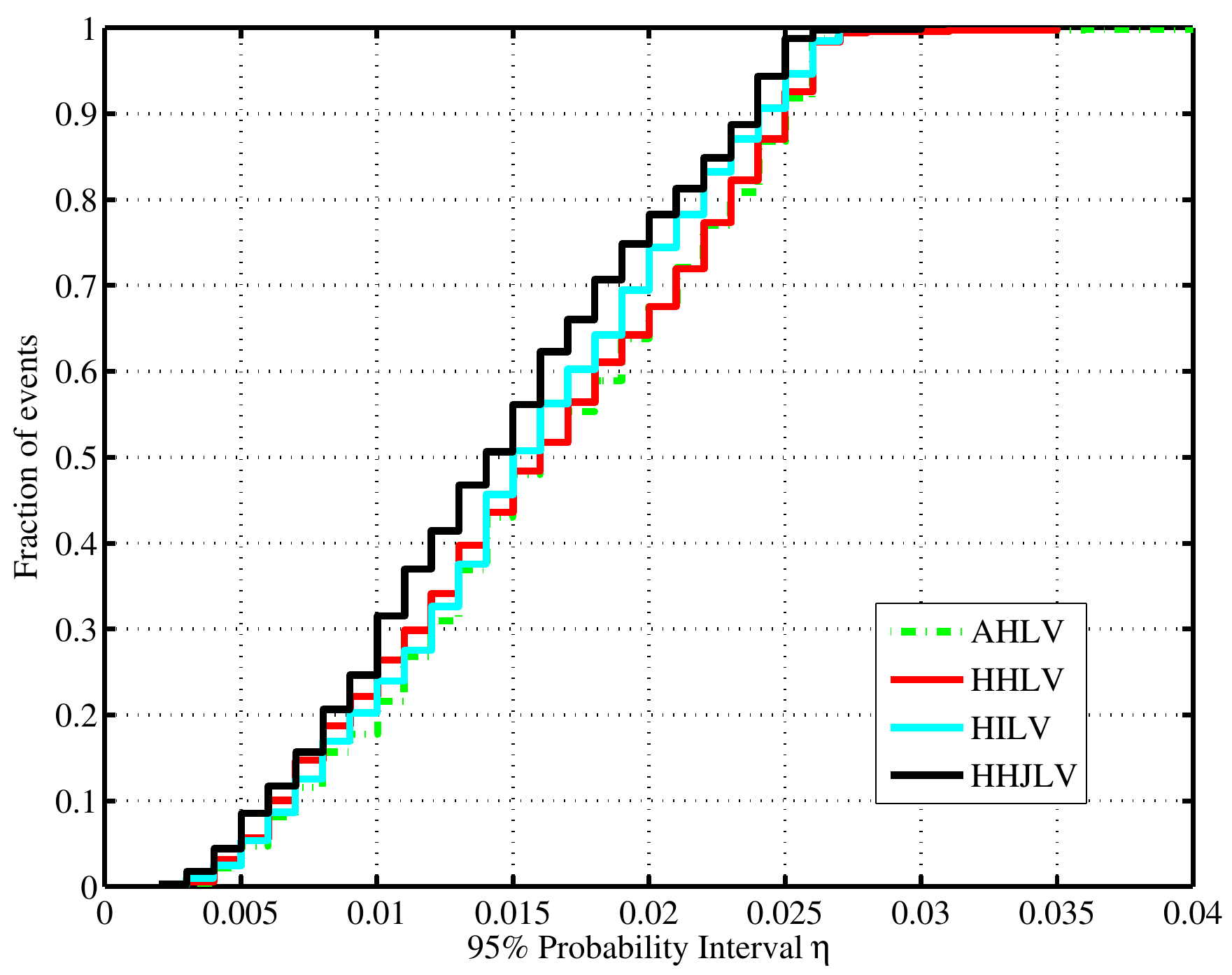}
\\
\includegraphics[width=\columnwidth]{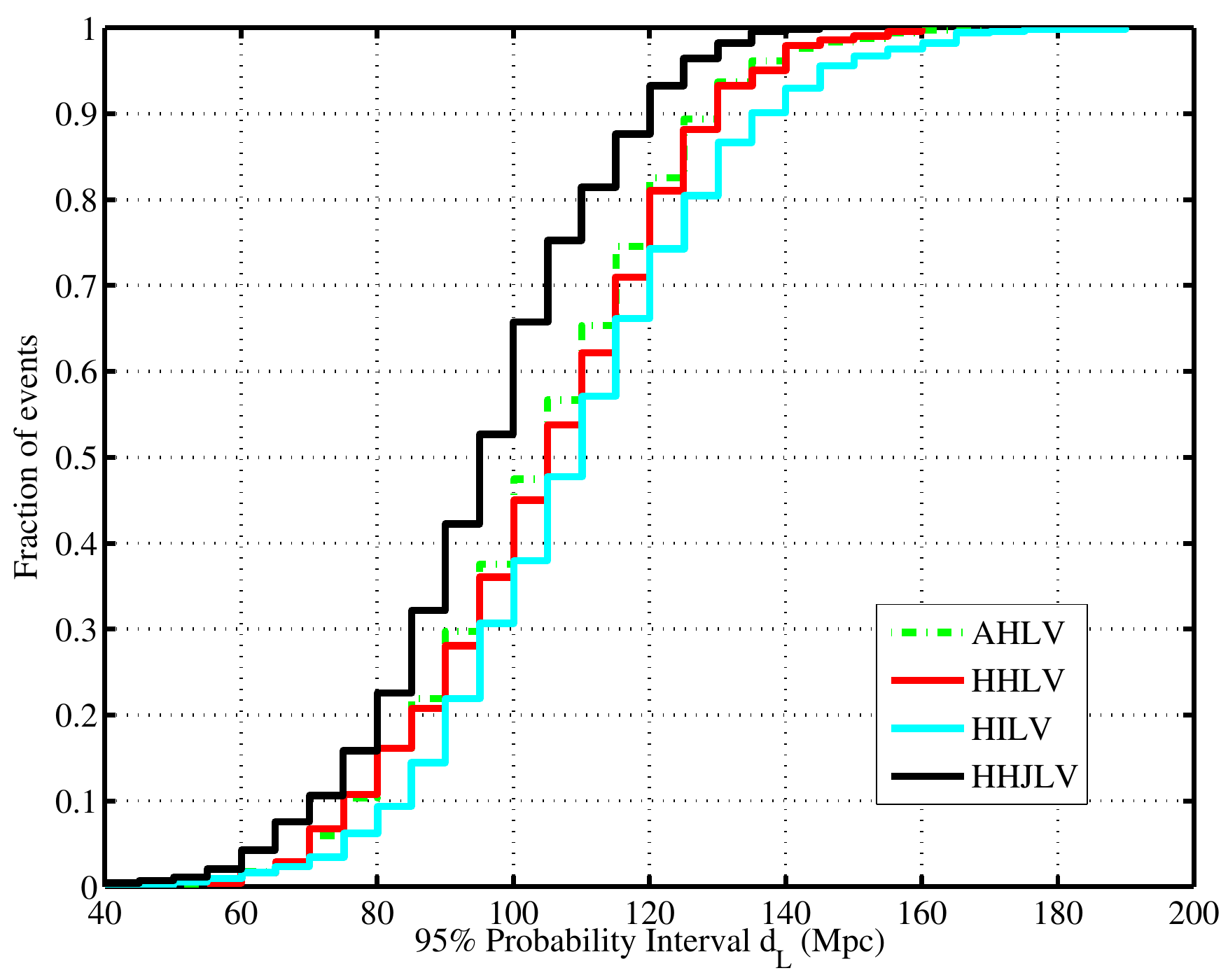}
\includegraphics[width=\columnwidth]{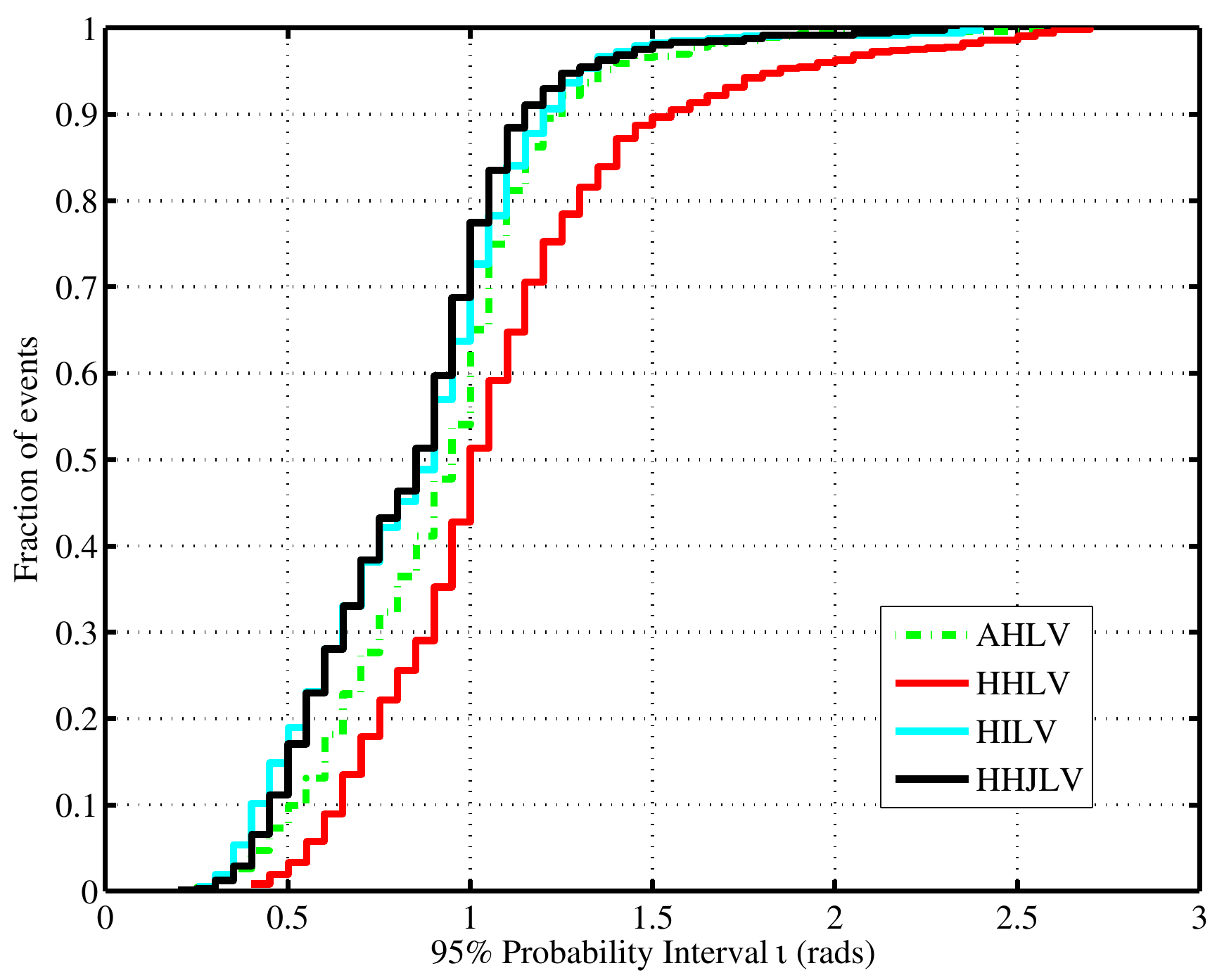}
\\
\includegraphics[width=\columnwidth]{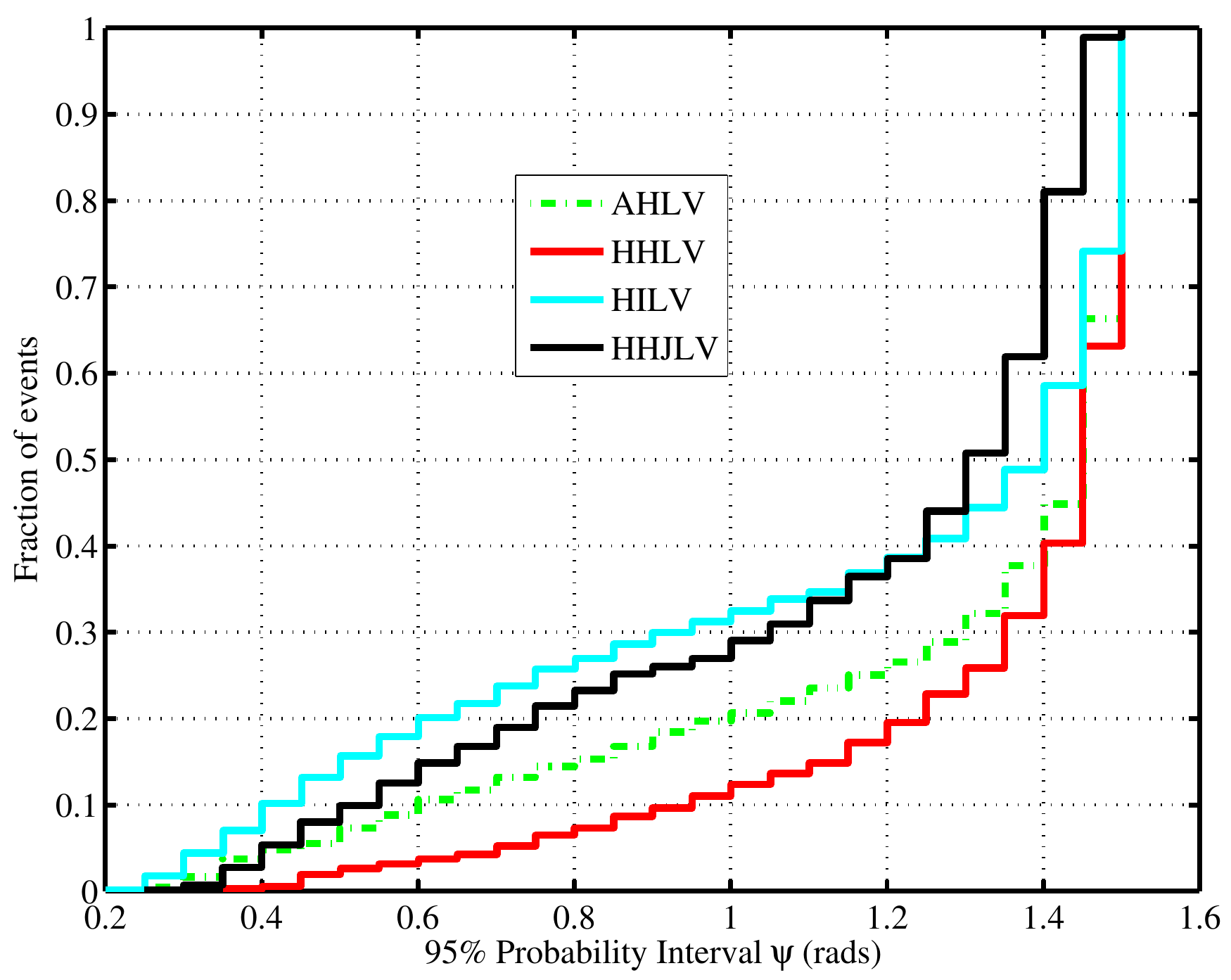}
\includegraphics[width=\columnwidth]{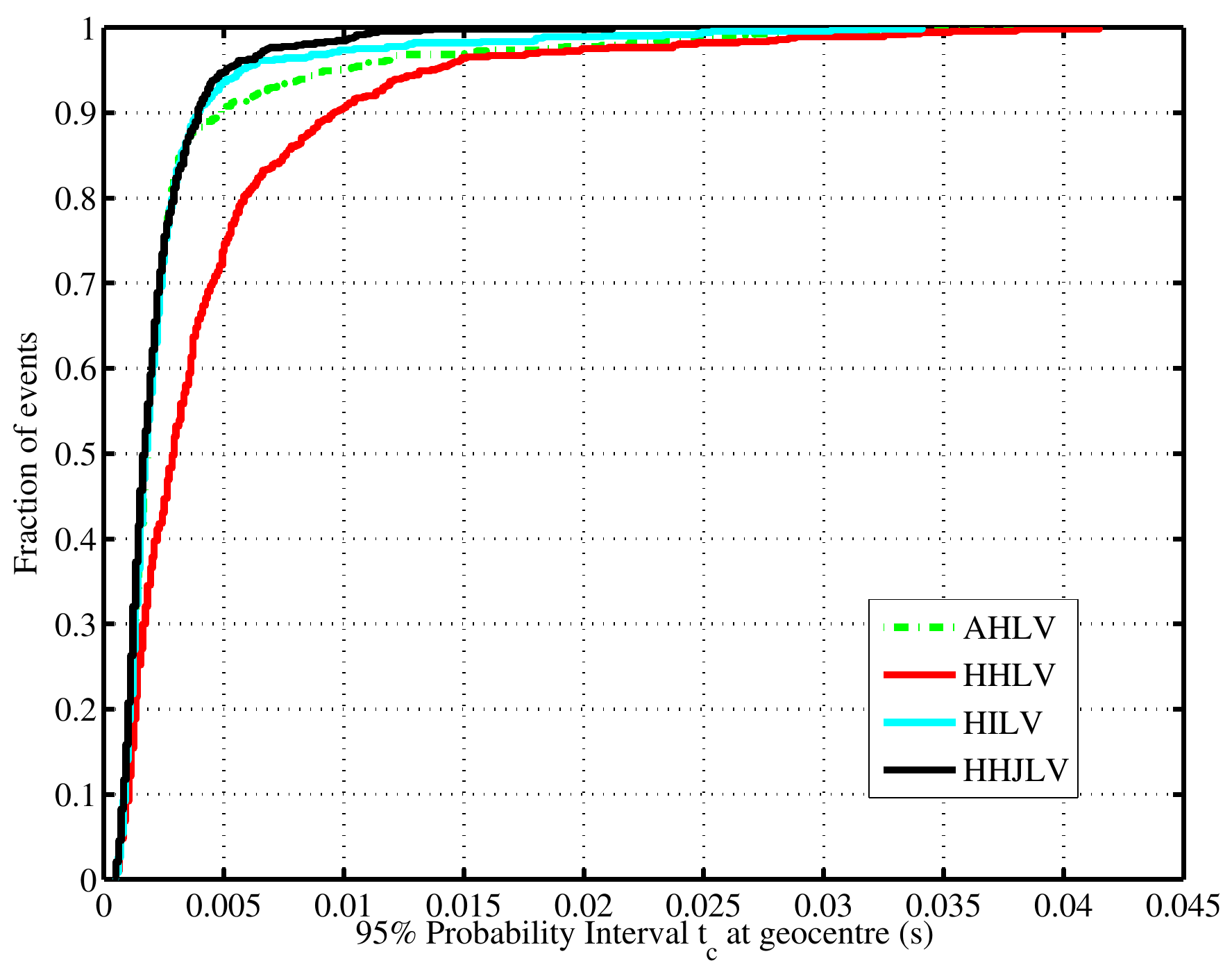}
\caption{\label{fig:NS_CDF}Cumulative histograms comparing the ability of the networks to resolve the individual parameters of main interest. Plots show fraction of events found within a given width of the confidence interval. The chirp mass $\Mc$ and symmetric mass ratio $\eta$ (top row) do not benefit significantly from a 4th site, except through better uniformity of SNR across the sky. The distance and inclination angle (middle row) show slight improvements in their resolvability but the effect is most marked in the 2D confidence intervals (see figure \ref{fig:dist_iota_cdf}). The polarization angle $\psi$ shows interesting behaviour, with the additional sites able to resolve a greater fraction of the signal to within a given interval of $\psi$, but with the median ratio of intervals being approximately unity (see table \ref{tab:inspnest}). This is explained by the sizeable fraction of signals where the polarization angle is not well resolved. The resolution of the time of coalescence parameter $t_c$ is similar for all 4-site networks, with 90\% of signals resolved within 5\,ms. The longer tail of less well resolved times agrees with figures \ref{fig:inspnest90pc} and \ref{fig:Fisher-sa} which show a hint of longer tail in sky resolution for AHLV compared to HILV.}
\end{figure*}

\begin{figure}
\includegraphics[width=\columnwidth]{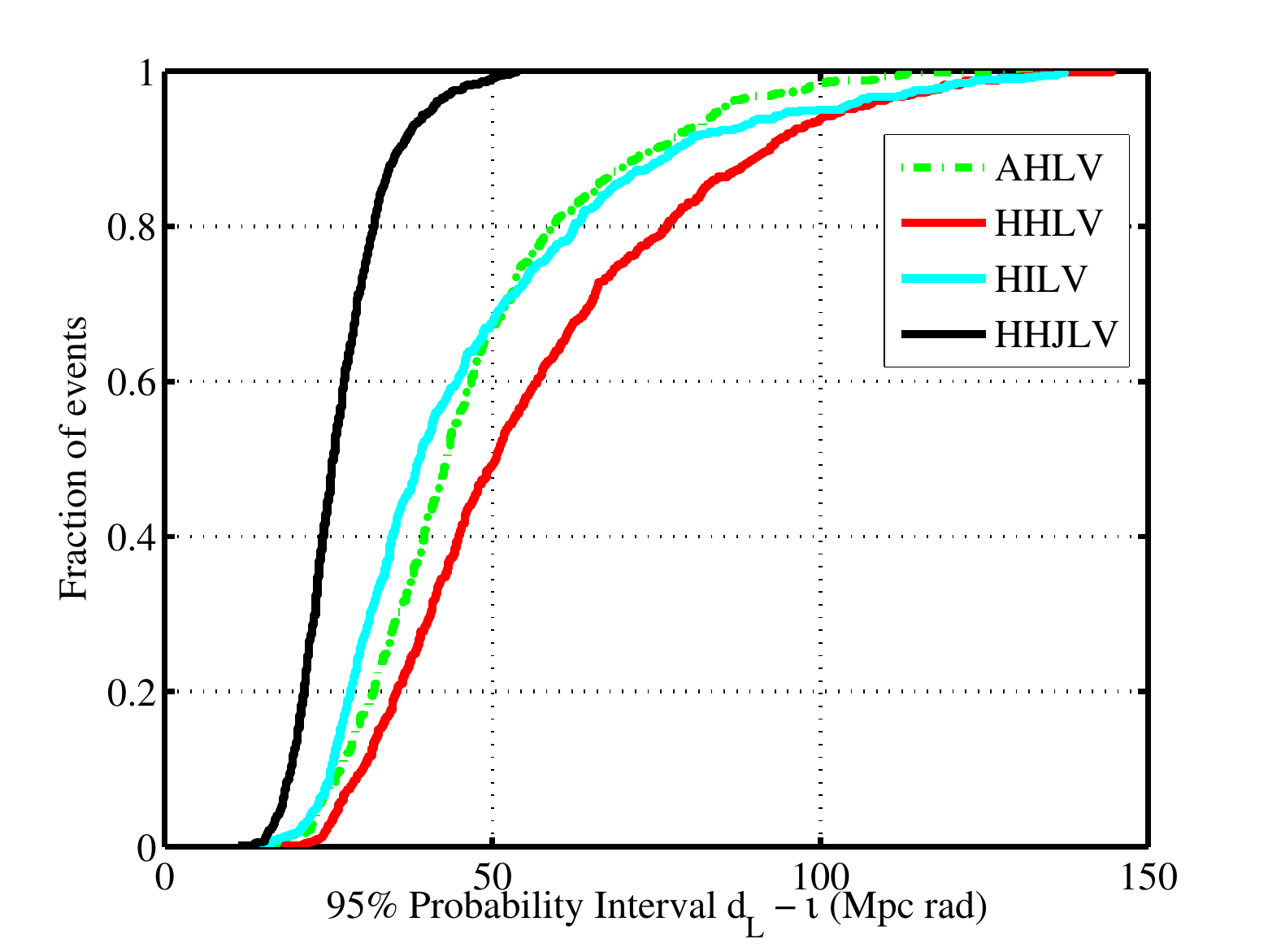}
\caption{\label{fig:dist_iota_cdf}Cumuative histogram comparing the ability of the networks to resolve the signal in 2D $d_L\text{--}\iota$ space. The improvement in parameter estimation with the 4-site networks due to elimination of the secondary mode is more striking here than in the individual $d_L$ and $\iota$ parameters (c.f. figure \ref{fig:NS_CDF}).}
\end{figure}

Both these effects are clearly visible in figure \ref{fig:degeneracy}, which shows the breaking of the degeneracy in $d_L\text{--}\iota\text{--}\alpha\text{--}\delta$ space as projected onto the $d_L\text{--}\iota$ and $\alpha-\delta$ planes.  In this example all the expanded networks allow the secondary maximum to be eliminated.
Figure \ref{fig:dist_iota_cdf} shows the cumulative distribution of the 95\% probability region for the $d_L\text{--}\iota$ space, with the improvement in two dimensions more prominent than looking at $d_L$ or $\iota$ individually (c.f. figure \ref{fig:NS_CDF}).

On the other hand, the accuracy with which mass parameters are measured does not improve as a consequence of moving the site of the fourth detector to India or Australia.  We can speculate that the reason for this is that information about masses comes primarily from the phase evolution of the signal, and the accuracy in mass estimation is predominantly set by the overall network SNR.  Meanwhile, masses do not strongly correlate with extrinsic parameters (with the exception of the time of coalescence), so their estimation is not significantly improved by better sky localization or inclination measurements.  

The results from the Fisher information matrix are in qualitative agreement with the two Bayesian approaches regarding the partial breaking of the distance/inclination degeneracy achieved by moving a detector to Australia or India (leading to marginal improvements in both parameters, see table \ref{tab:fisher}).  They also indicate that the accuracy with which masses can be measured is not affected by the network choice.

\section{Conclusions}
In this paper we studied the effect on parameter estimation of different networks of advanced detectors. We employed two different Bayesian techniques and the Fisher information matrix to estimate the accuracy of parameter recovery. We analysed a set of injections distributed in a grid in the extrinsic parameter space (without varying the mass and distance of injections) with the \Inspnest code, and verified the results with \LALInferenceMCMC. We performed a large scale Monte Carlo simulation using the Fisher matrix method with constant-SNR injections. We found consistent results between the three methods, pointing to significant gains in sky localization (typically by a factor of $\sim 3$---$4$) and modest gains in distance and inclination measurements with a network including a fourth site. We found that the 4-site networks are able to better resolve the polarisation angle of the source, in the cases where this is possible. We found no significant effect on mass measurements.

Comparing the different network configurations, we found, as expected, the strongest improvement in sky localization capability when the longest baseline (namely AHLV) was used, but that a site in India also provides a significant improvement in sky resolution. The HHJLV network, with the shortest extra baseline, provides the weakest improvement in sky resolution at a fixed signal-to-noise ratio; however, the fifth detector in this network can mitigate this, for an overall performance similar to HILV, but with fewer signals in the tail of the distribution with poor resolution.

We also find good agreement with previous work.  In particular, Fairhurst \cite{FairhurstAdvancedNetworks} finds 20-50\% of signals localised within 20\,$\mathrm{deg}^2$ for HHLV, and up to 20\% within 5\,$\mathrm{deg}^2$ with HHJLV, for an ensemble of sources at fixed distance of $160$\,Mpc, in good agreement with figure \ref{fig:inspnest90pc}. Despite the use of a different population of sources, Nissanke \emph{et al} \cite{Nissanke:2011} find results which qualitatively agree with our own. Comparison of the Fisher matrix results in figure \ref{fig:Fisher-sa} with Wen and Chen \cite{WenChen:2010} shows good qualitative agreement with their expected distribution for the HLV network at fixed SNR of 15, when taking into account a factor of $(30/15)^2=4$ for the difference in SNR used (30 in our case, 15 in theirs).

%We also found comparable behavior for two different orientations of the LIGO Australia detector, leading us to conclude that the orientation of its arms does not appreciably impact parameter-estimation accuracy.

In the present study, we focused on binary neutron stars (NS), which are the most ``confident'' source for the advanced detectors, but which are not expected to have significant spins \cite{MandelOShaughnessy:2010}.  On the other hand, black holes (BH) in NS-BH or BH-BH binaries can be rapidly spinning.  Previous studies (see, e.g., \cite{vanderSluys:2008b}) have shown that the presence of spin in one or both binary components can aid sky localization by providing additional polarization information through the precession inherent in misaligned spinning binaries.  Localization may be further enhanced when a signal from a spinning binary is captured by a four-detector network; on the other hand, improved resolution of extrinsic parameters with the help of a fourth detector site may aid in the reconstruction of astrophysically interesting quantities such as spin-orbit misalignment angles.

The improved ability to localize sources on the sky will be crucial in any search for electromagnetic counterparts to detected gravitational-wave signals (e.g., \cite{Bloom:2009,NuttallSutton:2010,Mandel:2011}).  Accurate measurements of the location of the merging binary can also be useful even in the absence of electromagnetic counterparts, for example, in measuring typical binary kick velocities \cite{Kelley:2010}.  We thus conclude that scientific considerations strongly favor an international gravitational wave network with four or more sites.

\section*{Acknowledgements}

The authors would like to thank B. S. Sathyprakash, Stephen Fairhurst, Samaya Nissanke, and Jonathan Gair for useful discussions, and Kazuaki Kuroda for providing the LCGT noise curve. JV and BA were supported by the Science and Technology Facilities Council. IM was partially supported by the NSF Astronomy and Astrophysics Postdoctoral Fellowship, award AST-0901985; IM is also grateful for the hospitality of the Aspen Center for Physics during the end stages of preparing this paper.  BF acknowledges the NSF GK-12 grant, award DGE-0948017. VR and VK acknowledge NSF grant PHY-0969820.  CR acknowledges a Northwestern University graduate fellowship. AV was partially supported by the Science and Technology Facilities Council of the United Kingdom

\appendix*
\section{False alarm probabilities in a coherent search}

It is instructive to ask how different network configurations affect false alarm probabilities (FAPs), or, alternatively, what the detection thresholds would need to be for a fixed FAP.  We consider only the case of a fully coherent search; the FAP of triggers in searches that only require coincidence between detectors in mass and time of coalescence parameters, such as \cite{S5LowMassLV,Collaboration:S5HighMass}, should not depend on the accuracy of measuring the sky location and other extrinsic parameters.  If a coherent search is used, and all detectors are assumed to have independent noise, we might expect that the coherence constraint would be stronger for two co-located detectors that should have the same signal, and therefore HHLV should have a lower FAP than AHLV or HILV for a fixed network SNR. 
%Bayesian methods provide a simple way to quantitatively estimate this effect.

We provide a Bayesian treatment of this question by comparing the odds ratio between the coherent signal hypothesis $GW$ and the noise hypothesis $N$ for different network configurations.
The odds ratio is just the ratio of posterior probabilities for the two models,
\begin{equation}
B_{GW,N} = \frac{P(GW|d)}{P(N|d)}=\frac{P(GW)}{P(N)}\frac{P(d|GW)}{P(d|N)}.
\end{equation}
The only term that depends on the network configuration is the evidence for the presence of a signal, 
$Z_{GW} \equiv P(d|GW)$, which can be written as
%As the prior odds $P(GW)/P(N)$ of a gravitational wave being present cannot be affected by moving one of the detectors, and we are assuming the fourth detector produces Gaussian noise with the same PSD in both cases, so $P(d|N)$ does not change, any change in odds must come from the evidence for a signal $P(d|GW)=Z_{GW}$ changing.
%The evidence $Z_{GW}$ can be written as
\begin{equation}
Z_{GW}=\int{}p(\vec{\theta}|GW)p(d|\vec{\theta},GW)d\vec{\theta}.
\end{equation}
We assume for the sake of simplicity that the prior $p(\vec{\theta}|GW)=k$ is constant in the small region where the likelihood is significant, and the likelihood $p(d|\vec{\theta},GW)$ is Gaussian in the model parameters about a maximum $L_\mathrm{max}$ at $\vec{\theta}_0$ (as assumed for the Fisher matrix calculation), 
\begin{equation}
p(d|\vec{\theta},GW) \approx L_\mathrm{max}\exp{\left[-\frac{1}{2}({\vec{\theta}-\vec{\theta}_0}){\bf C}^{-1}({\vec{\theta}-\vec{\theta}_0})^T\right]}.
\end{equation}
This yields
\begin{equation}
\label{eqn:approxZ}
Z_{GW} = k L_\mathrm{max}(2\pi)^{-N/2}\sqrt{\det{\bf C}},
\end{equation}
where ${\bf C}$ is the covariance matrix of the parameters, $N$ is the dimensionality of the model, and $L_\mathrm{max}$ is the maximum likelihood of the data, given by
\begin{equation}
L_\mathrm{max}\propto \exp\left({-\frac{1}{2}\left<\tilde{d}-\tilde{h}(\vec{\theta}_0)|\tilde{d}-\tilde{h}(\vec{\theta}_0\right>}\right) = \frac{\rho^2}{2}.
\end{equation}
For a fixed network SNR (maximum likelihood), the evidence is therefore proportional only to $\sqrt{\det{\bf C}}$, which scales with the size of the region in parameter space to which a signal's parameters can be constrained. 
%Assuming that at the maximum likelihood, the data $\tilde{d}=\tilde{h}(\vec{\theta}_0)+n$ consists of the signal corresponding to the waveform with the maximum likelihood parameters and Gaussian noise, then the inner product is $\left<\tilde{d}-\tilde{h}(\vec{\theta}_0)|\tilde{d}-\tilde{h}(\vec{\theta}_0\right>=\left<n|n\right>$, which is a constant between network configurations if the noise realizations are the same. The evidence $Z$ is therefore proportional only to $\sqrt{\det{\bf C}}$, which scales with the size of the region in parameter space to which a signal's parameters can be constrained. 

Thus, for a fixed SNR, the Bayes factor is larger when parameters are less precisely estimated.  Due to the shrinking size of the posterior distribution in sky location, distance and inclination, we might expect a decrease by a factor of $\sim 5$ in the allowed fraction of the prior volume between the HHLV and AHLV  or HILV networks. Then, for a fixed detection threshold $\rho_{\rm min}$, the odds ratio will be a factor $\sim 5$ smaller for the AHLV or HILV network than for the HHLV network. Conversely, if we want to keep the same false alarm probability (same minimal odds ratio required for detection), the SNR threshold $\rho_{\rm min}^2 \sim 2 \log{L_\mathrm{max}}$ for the AHLV or HILV network must increase by $2\log{5}$ relative to the HHLV network.  This corresponds to a very modest increase in the SNR: for example, if the threshold is $12$ for the HHLV network, it will only rise to $12.13$ for the AHLV or HILV network.

The analysis above assumes that the noise in all interferometers is uncorrelated.  While that is a reasonable assumption for distant interferometers, various environmental factors can lead to correlated noise in co-located interferometers, such as the two possible detectors at Hanford.  This correlation can increase the threshold necessary for detection.  For example, at times when only two of the four detectors in our presumed network are operational, it may not be possible to make a convincing detection if only the two Hanford detectors are operating from the HHLV network, while any two detectors from the AHLV or HILV network have a chance to detect a sufficiently loud signal.

\bibliography{Bibliography}{}
\end{document}